\documentclass[lettersize,journal]{IEEEtran}

\usepackage{amsmath,amssymb,amsfonts}
\usepackage{algorithmic}
\usepackage{algorithm}
\usepackage{array}
\usepackage{multirow}
\usepackage{subfig}
\usepackage{textcomp}
\usepackage{stfloats}
\usepackage{url}
\usepackage{verbatim}
\usepackage{graphicx}
\usepackage{cite}
\usepackage{booktabs}
\usepackage{hyperref}
\usepackage[table, dvipsnames]{xcolor}
\usepackage{caption}
\usepackage{subcaption}
\usepackage{todonotes}
\usepackage{orcidlink}

\usepackage{minted}
\usepackage{tabularx}
\usepackage[table]{xcolor}
\usepackage{tcolorbox}
\usepackage{placeins}

\definecolor{reviewercolor}{RGB}{245, 245, 245}
\definecolor{revisedblue}{RGB}{0, 0, 200}

\newcommand{\changed}[1]{#1}

\newcommand{\implication}[1]{%
    \begin{center}
        \fbox{%
            \begin{minipage}{0.95\linewidth}
                \textbf{Implication}: #1
            \end{minipage}%
        }
    \end{center}%
}

\title{Multi-Dimensional Autoscaling of Stream Processing Services on Edge Devices}

\author{%
Boris Sedlak~\orcidlink{0009-0001-2365-8265},~\IEEEmembership{Member,~IEEE},\
Philipp Raith~\orcidlink{0000-0003-3293-9437},~\IEEEmembership{Member,~IEEE},\
Andrea Morichetta~\orcidlink{0000-0003-3765-3067},~\IEEEmembership{Member,~IEEE},\
V\'ictor Casamayor Pujol~\orcidlink{0000-0003-2830-8368},~\IEEEmembership{Member,~IEEE},\
Schahram Dustdar~\orcidlink{0000-0001-6872-8821},~\IEEEmembership{Fellow,~IEEE}%
\thanks{B. Sedlak, P. Raith, A. Morichetta, and S. Dustdar are with the Distributed Systems Group, TU Wien, Vienna, Austria (e-mail: b.sedlak@dsg.tuwien.ac.at; p.raith@dsg.tuwien.ac.at; a.morichetta@dsg.tuwien.ac.at; dustdar@dsg.tuwien.ac.at).}%
\thanks{V. Casamayor Pujol and S. Dustdar are with the Department of Engineering, Universitat Pompeu Fabra, Barcelona, Spain and Dustdar is with ICREA (e-mail: v.casamayor@upf.edu; schahram.dustdar@upf.edu).}%
}

\IEEEpubid{0000--0000/00\$00.00~\copyright~2025 IEEE}
\IEEEpubidadjcol

\usepackage{flushend} 

\begin{document}

\maketitle

\begin{abstract}
Edge devices have limited resources, which inevitably leads to situations where stream processing services cannot satisfy their needs. While existing autoscaling mechanisms focus entirely on resource scaling, Edge devices require alternative ways to sustain the Service Level Objectives (SLOs) of competing services.
To address these issues, we introduce a Multi-dimensional Autoscaling Platform (MUDAP) that supports fine-grained vertical scaling across both service- and resource-level dimensions. MUDAP supports service-specific scaling tailored to available parameters, e.g., scale data quality or model size for a particular service. To optimize the execution across services, we present a scaling agent based on Regression Analysis of Structural Knowledge (RASK). The RASK agent efficiently explores the solution space and learns a continuous regression model of the processing environment for inferring optimal scaling actions. We compared our approach with two autoscalers---the Kubernetes VPA and a reinforcement learning agent---for scaling up to 9 services on a single Edge device. Our results showed that RASK can infer an accurate regression model in merely 20 iterations (i.e., observe 200s of processing). By increasingly adding elasticity dimensions, 
RASK sustained the highest request load with 28\% fewer SLO violations, compared to baselines.
\end{abstract}



\section{Introduction}
\label{sec:introduction}

The Edge layer has become a pillar~\cite{de2019foundations} for services that demand low-latency, high-reliability access to computing, such as in  automotive~\cite{liu2021vehicular} or disaster response~\cite{dazzi2024urgent} scenarios.
In such dynamic and ever-changing scenarios, optimizing the resource allocation and ensuring performance guarantees---quantified through Service Level Objectives (SLOs)---is paramount. To assign processing services the desired resources under changing conditions (e.g., fluctuating demand), elastic computing principles~\cite{dustdar2011principles} and autoscaling policies provide a remedy.
In this context, the Kubernetes Vertical Pod Autoscaler (VPA), as introduced in 2023~\cite{kubescaler2023vpa}, offers fast resource adaptation and service elasticity---essential for resource and time-sensitive scenarios. \changed{Albeit serverless computing emerges as a solution for flexible application management, it still suffers from performance and resource management issues~\cite{li2022serverless,gajanin2025performance}, especially during request peaks. 
Hence, using dependable, lightweight containers~\cite{taleb2025survey} remains the most reliable solution.}

To ensure high-level SLOs, like efficiency or accuracy~\cite{nastic_sloc_2020}, applications need to adjust and optimize a myriad of lower-level components.
\changed{Yet, traditional autoscalers, including the Kubernetes VPA, have two major limitations:
Firstly, they often use simple rule-based decision-making, and its effectiveness highly depends on the operator to manually fine-tune it.
Therefore, researchers and practitioners have advocated learning-based solutions---prominently using Reinforcement Learning (RL)~\cite{qiu2020firm,xue2022meta,qiu2023aware,wang2024deepscaling,mayerhofer2025hpaqt,nayir2025multi,xu_coscal_2022}. Still, these approaches come with other issues, including long, sample-inefficient training periods, and limited interpretability of inferred actions.
Secondly, to fulfill SLOs, traditional autoscalers are limited to claiming additional resources---hence, they operate only in one \textit{elasticity dimension}.}
While this simplifies service orchestration, it does not exploit the manifold of potential elasticity strategies across different services. In particular, video inference services might dynamically scale the service quality~\cite{furst_elastic_2018} or model size~\cite{romero_infaas_2021}; despite the huge impact of these parameters on the service throughput, they are not commonly used in autoscaling. 
\changed{Notably, individual Edge devices can only provide resources up to a strict limit. To ensure processing SLOs regardless, computation is commonly redirected through offloading~\cite{sedlak_slo-aware_2024} or horizontal autoscaling~\cite{quattrocchi_autoscaling_2024_short}.
However, emerging architectures like the Computing Continuum~\cite{cardellini_scalable_2025} face volatile networking and resource availability; hence, they need additional, decentralized mechanisms to optimize SLO fulfillment. Thus, in this paper, we assume the complete absence of remote resources that support offloading or horizontal scaling.}

\changed{Our work addresses this gap by offering a flexible, \textit{multi-dimensional} autoscaling platform---controlling both the infrastructure and the application~\cite{xu2025auto}---for harmonizing the SLO fulfillment on resource-constrained Edge devices.}
Conceptually, we first analyze the behavior and SLO fulfillment of all deployed services through a regression model, and then optimize the device-wide SLO fulfillment through a numerical solver.
The numerical solver provides fine-grained assignments for vertically scaling both service- and resource-level parameters within an Edge node.
By adding the control at the device level, we ensure that scaling actions are harmonious and do not penalize other services.
We implement our solution \textit{service-agnostic}, making it modular for adding new service types or elasticity parameters. Likewise, our methodology can cope with changing load patterns or SLO thresholds and ensure its model accuracy despite distribution shifts.

\captionsetup[subfloat]{justification=centering}
\begin{figure*}[t]
    \centering
    \subfloat[An IoT stream is processed at the closest Edge device. The processing service has enough resources allocated to fulfill its two SLOs.]{
        \includegraphics[width=0.304\textwidth]{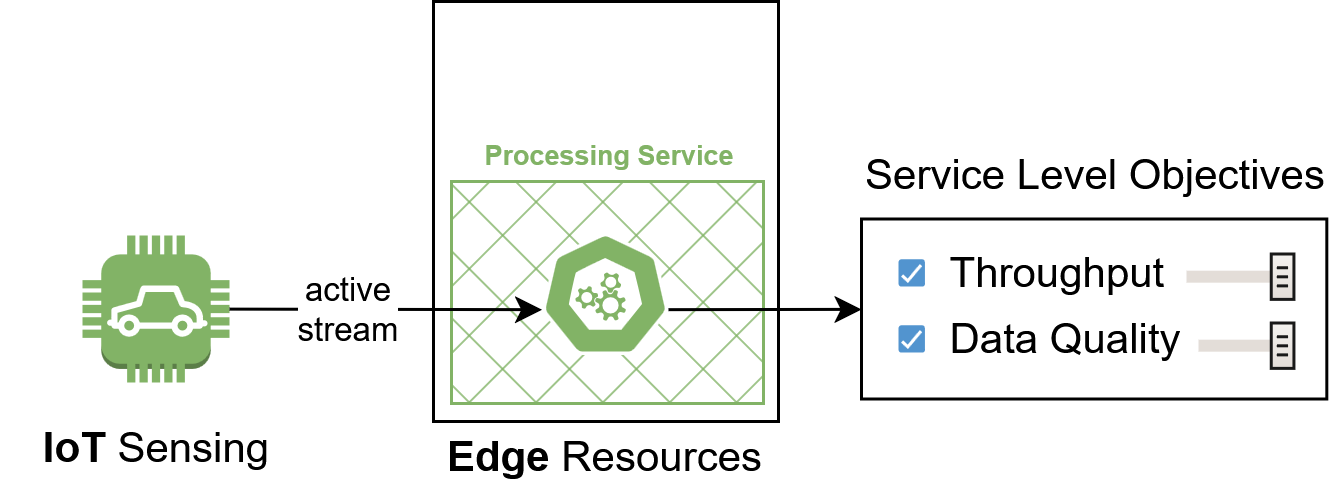}
        \label{fig:scenario-1}
    } \hfill
    \subfloat[Another IoT device wishes to process sensor data at the Edge device. However, to fulfill its SLOs, it requires more resources than available.]{
        \includegraphics[width=0.319\textwidth]{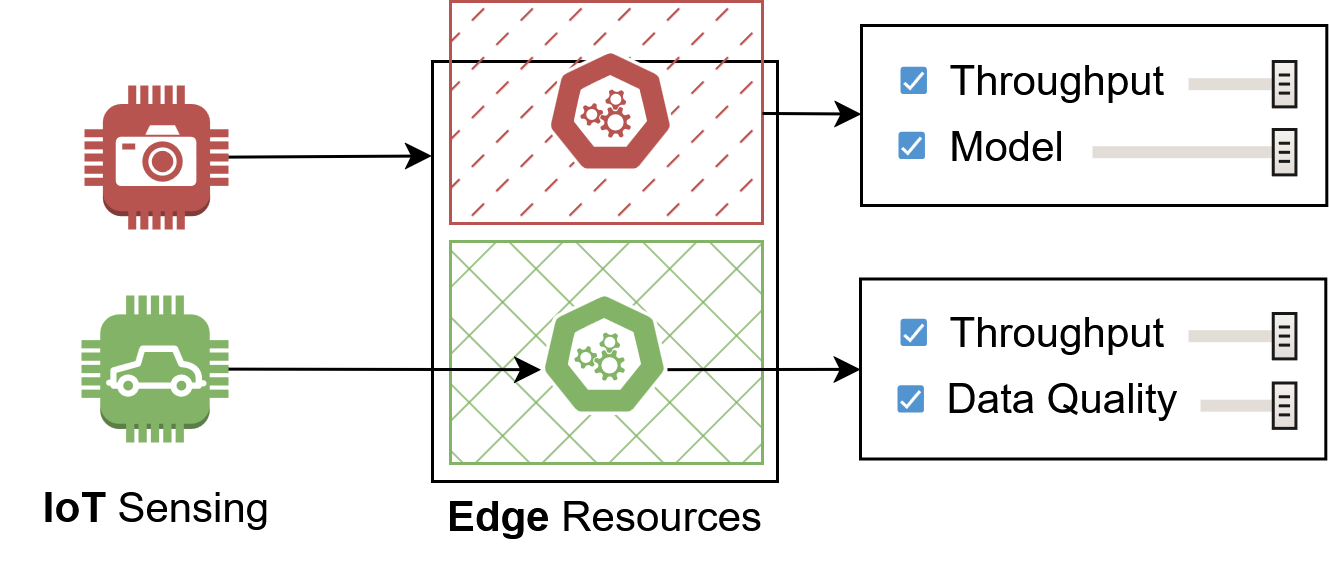}
        \label{fig:scenario-2}
    } \hfill
    \subfloat[To accommodate more processing services at the Edge device, the services can decrease the provided quality and the respective resource demand.]{
        \includegraphics[width=0.329\textwidth]{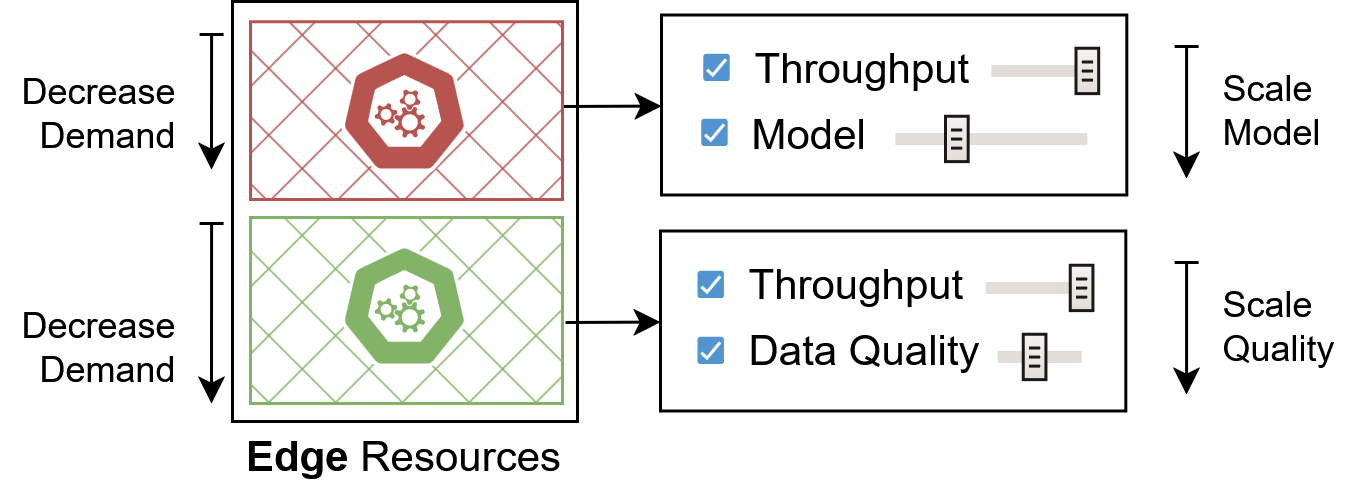}
        \label{fig:scenario-3-1}
    }
    \caption{Co-locating multiple processing services on a constrained Edge device by accurately trading off service quality.}
    \label{fig:combined-scenarios}
\end{figure*}
\captionsetup[subfloat]{justification=justified}

\IEEEpubidadjcol

An effective way to illustrate the impact of our approach is through a key scenario in Fig.~\ref{fig:combined-scenarios}: Sensor data from different sources (vehicles, cameras, etc.) is processed by services on an Edge device. This is a time-sensitive, dynamic environment where suboptimal decisions can lead to dangerous outcomes. 
However, when multiple services have to operate on the same node~(see~Fig.~\ref{fig:scenario-2}), it is not possible to meet their resource demands, creating conflicts and mitigating performance. In this case, downscaling the allocated resources for an object detection model can impact its reactivity or, in the worst case, break its execution. 
For this reason, our approach enables multi-dimensional elasticity control over services. This solution allows trading off qualitative aspects, like data quality or model size, for sustaining essential functionality, like throughput. \changed{This approach---also known as \textit{brownout}~\cite{klein2014brownout}---is downgrading the service quality to avoid resource saturation. In our case, we tailor scaling actions to the individual characteristics of processing services, enabling optimized configurations and SLO fulfillment across both resource and service dimensions.} At the same time, our solver ensures fair resource allocation to guarantee the performance of all concurrent services. 
Overall, we contribute to advance the state-of-the-art through:

\begin{itemize}
    \item MUDAP, a Multi-dimensional Autoscaling Platform for resource-constrained Edge environments that supports fine-grained vertical scaling of elasticity parameters \changed{without the need to restart processing containers}. It allows tailor-made scaling for heterogeneous services by exposing particular service- or resource parameters. 
    %
    
    \item RASK, a regression-based scaling agent that optimizes the SLO fulfillment across multiple competing processing services. RASK creates an explainable regression model that allows it to infer optimal scaling decisions.
    
    \item An extensive evaluation that highlights superior SLO fulfillment of multi-dimensional autoscaling under dynamic conditions. We compared RASK with existing autoscalers, including the Kubernetes VPA, and demonstrated how RASK sustains periods of high load with 28\% fewer SLO violations, while barely introducing any CPU overhead. To achieve this performance, RASK was extremely sample-efficient, requiring merely 20 training iterations---corresponding to 200s of processing.
\end{itemize}

The remainder of the paper is structured as follows. Section~\ref{sec:preliminaries} introduces the core challenges and background of this work. Section~\ref{sec:platform} presents our MUDAP architecture and its main components, whereas in Section~\ref{sec:agent} we introduce our RASK agent. Section~\ref{sec:evaluation} offers an in-depth evaluation of our proposed approach in comparison to reference baselines. Finally, in Section~\ref{sec:discussion} we outline the implications of our work, while Section~\ref{sec:conclusion} concludes the paper.

\section{Preliminaries}
\label{sec:preliminaries}


\subsection{Motivating Example}

Consider a smart city that consists of numerous IoT sensors and processing infrastructure distributed throughout a wide area. To optimize traffic flow and reduce emissions, we want to process various types of data streams (e.g., audio, video, temperature) through a series of containerized services. Consider Fig.~\ref{fig:MUDAP-architecture}, where an IoT device, like an IP camera, generates a continuous stream of video frames. To detect objects within the stream, e.g., cars in traffic junctions, video frames are processed by an Edge device in close vicinity.

\changed{Frames are collected in a standalone buffer, from where services consume them;} this provides two benefits: (1) new data sources, like another camera, can simply stream to the same buffer; and (2) if the amount of generated data increases, a processing service can react to this according to the backpressure. From a service perspective, these changes only affect the buffer size. Therefore, we abstract the actual data sources as \textit{requests per second} (RPS) received by the buffer. To sustain SLO fulfillment under varying RPS, services are scaled dynamically. \changed{While conventional scaling mechanisms, like horizontal autoscaling~\cite{quattrocchi_autoscaling_2024_short}, assume that tasks can be shared among machines, we constrain the problem further and assume an isolated Edge device with limited resources.}

\subsection{Core Challenges}

Considering this example, we formulate three challenges that address (\textit{C1}) SLO fulfillment on resource-constrained devices, (\textit{C2}) multi-dimensional autoscaling for Edge devices and (\textit{C3}) achieving SLO targets under dynamic workloads.

\begin{itemize}
    \item \textbf{Challenge 1}: Vertical scaling can increase resource efficiency and SLO fulfillment~\cite{wang2020dyverse}. However, scaling multiple services on an Edge device is challenging because services share limited resources. Typical approaches to vertical scaling, such as CPUs, thus have limited impact for improving service throughput. Service-aware vertical scaling, such as input size, could mitigate this issue and trade off quality in exchange for improved throughput.

    \vspace{2pt}
    \item \textbf{Challenge 2}: Vertical scaling approaches must be tailored towards services, their particular resource needs, and the processing device. The impact of vertical scaling differs across services: while an additional CPU core or lower data quality might increase the throughput of one service, it might not affect another service type at all. Learning the impact of scaling actions per service is challenging and exacerbated in increasingly larger action spaces.
    
    \vspace{2pt}
    \item \textbf{Challenge 3}: Fluctuating usage pattern mean the amount of IoT data---or IoT data sources---changes throughout the day. To keep SLOs fulfilled, while avoiding the risk of overprovisioning resources, the allocated resources must be adjusted. This is especially important in scenarios where multiple services on an edge device receive bursts of requests at the same time, calling for actions that optimize latency or throughput with resources at hand.
\end{itemize}

\subsection{Background \& Related Work}

In the following, we describe two concepts that are central to our work and processing in general: SLOs and elasticity strategies. We reflect on their current state-of-the-art and present how they are used and extended in this work.

\vspace{3pt}
\subsubsection{Service Level Objectives}


In their classical sense~\cite{sharma_sla_2023,keller_wsla_2003}, SLOs track basic functional service aspects like availability or latency.
%
In their wider sense, SLOs are increasingly used to track composed metrics that express high-level goals like effectiveness~\cite{nastic_sloc_2020} or accuracy~\cite{danilenka_adaptive_2024}---possibly supported through an SLO language~\cite{pusztai_slo_2021} or orchestration framework~\cite{pusztai_polaris_2022}. While these approaches are arguably still immature, they allow consumers to freely express requirements, rather than choose between predefined SLOs.
Hence, in the context of this paper, we adopt the \textit{wider sense of SLOs}~\cite{sedlak_controlling_2023} for tracking functional aspects (e.g., latency or throughput) and non-functional aspects (e.g., video quality or model accuracy) alike. 

Generally, an SLO $q$ relates a variable to a target value $t_q$; for example, keeping service throughput ($tp$) $\geq 30$.
Given a metrics ($m \in M$) and an SLOs ($q \in Q$), we calculate the SLO fulfillment ($\phi$)\footnotemark as a continuous value through
%
\begin{equation}
\phi(q, m) = 
\begin{cases}
\frac{m}{t_q} & \text{if } m < t_q \\
1.0 & \text{if } m \geq t_q
\end{cases}
\label{eq:slo-f}
\end{equation}
Under this representation, SLOs cannot be overfulfilled; hence, two metrics $m_\textit{tp} = 40$ and $m_\textit{tp} = 100$ would both achieve the maximum SLO fulfillment of $\phi = 1.0$.

\footnotetext{We choose the letter '$\phi$' due to its sound: SLO ful-\textit{phi}-llment}

\vspace{3pt}
\subsubsection{Elasticity Strategies}


The emergence of Cloud computing~\cite{buyya_cloud_2009} supported dynamic changes to provisioned resources. Autoscaling platforms, like Kubernetes (k8s), use this for horizontal and vertical autoscaling. \changed{This dimension for flexible resource management, known as \textit{elasticity}, has led to advanced control strategies. Recent work used this extensively for building RL-based solutions~\cite{qiu2020firm,xue2022meta,qiu2023aware,wang2024deepscaling,mayerhofer2025hpaqt,nayir2025multi,xu_coscal_2022}: given a few boundaries, the algorithm can learn the best strategy to dynamically adapt to application or infrastructure changes. Still, RL requires extensive training, accurate fine-tuning, and struggles with high-dimensional environments; hence, its use in production~\cite{xu2025auto,mayerhofer2025hpaqt} is still limited.}
%
Furthermore, while elasticity---as defined by~\cite{dustdar_principles_2011}---is not limited to resource scaling, scaling other dimensions (e.g., quality) has not found much traction. We hypothesize that the Cloud's abundance of resources made it obsolete to compromise quality. This led to a situation where contemporary research~\cite{chen_survey_2018,gari_reinforcement_2021,quattrocchi_autoscaling_2024_short,wang_dyverse_2020} on autoscaling focuses almost exclusively on adjusting or scheduling resources. \changed{Even prominent multi-dimensional approaches, like AWARE~\cite{qiu2023aware}, focus solely on scaling resource demands---although they combine vertical \textit{and} horizontal autoscaling.}

The shift toward Edge computing exposes the limitations of traditional orchestration \cite{kimovski_cloud_2021, petcu_service_2021} in managing constrained, heterogeneous resources. While collaboration strategies like offloading \cite{han_tailored_2021, wu_intelligent_2023} and the Computing Continuum \cite{cardellini_scalable_2025, spring_mach_2025} provide relief, they primarily displace the problem by assuming remote availability. To ensure SLO fulfillment in truly isolated or resource-scarce environments, scaling the service quality becomes the last remedy.



\changed{While there exist works that provide elastic service quality, they are often named differently: model-less inference~\cite{romero_infaas_2021} dynamically chooses the size of ML models according to SLOs and available hardware; other options to ensure SLOs are filtering the video content~\cite{sun_elasticedge_2022} or changing stream parameters~\cite{furst_elastic_2018,sedlak_equilibrium_2024}---like video resolution or frame rate. However, these works, in turn, do not support resource scaling.}

While recent work explores simultaneous scaling of resources and quality~\cite{sedlak_towards_2025_short,laso_multidimensional_2025}, these approaches remain impractical for several reasons. First, they tightly couple elasticity strategies with autoscalers---thus lacking a modular mechanism to register service-specific strategies or custom autoscaling algorithms. Second, selecting among elasticity strategies introduces an optimization problem that is addressed using model-free reinforcement learning. Such methods (e.g., Deep Q Networks) typically require thousands of iterations to converge. However, autoscaling actions commonly require multiple seconds to show effects~\cite{sedlak_towards_2025_short}, slowing down every single iteration of the agent. Finally, these approaches operate in a coarse-grained, discrete action space, which can prevent the system from reaching the global optimum.

Given that, we conclude that there is a clear gap for (1) an extensible multi-dimensional scaling platform that combines quality and resource scaling, while presenting clear interfaces for coupling different autoscalers; also (2) multi-dimensional autoscalers have not applied sample-efficient learning that allow (3) fine-grained adjustments of service-specific parameters. In the following, we address this through our modular autoscaling platform and regression-based scaling agent.

\section{Multi-dimensional Autoscaling Platform}
\label{sec:platform}

\begin{figure}
    \centering
    \includegraphics[width=1.0\columnwidth]{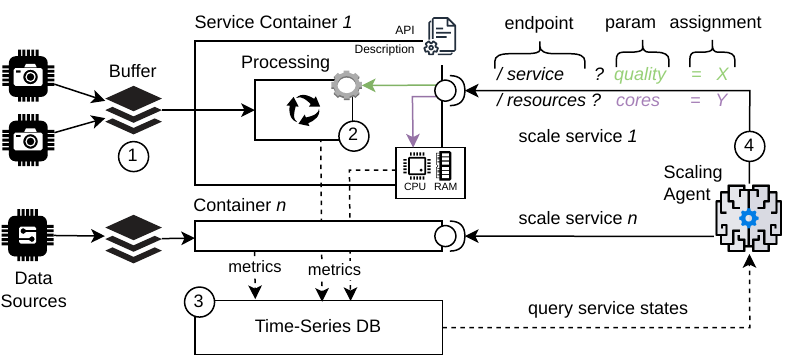}
    \caption{Conceptual architecture of the MUDAP platform: \textcircled{\small 1} IoT devices continuously ingest data into a buffer; \textcircled{\small 2} data is collected and processed every second; \textcircled{\small 3} container and service metrics are scraped by a time-series DB; \textcircled{\small 4} the scaling agent queries the service states and adjusts the service execution and resource limits through the exposed API endpoints.}
    \label{fig:MUDAP-architecture}
\end{figure}

Common autoscaling platforms are limited to resource scaling and hence, do not offer interfaces for coupling multi-dimensional scaling agents.
To that extent, this section describes a \underline{Mu}lti-\underline{d}imensional \underline{A}utoscaling \underline{P}latform (MUDAP) that exposes an API for dynamically scaling both service and resource parameters of containerized processing services in an Edge device. In Fig.~\ref{fig:MUDAP-architecture} we summarize the conceptual architecture of MUDAP, which we explain in the following.
Afterwards, in Section~\ref{sec:agent}, we present a scaling agent that uses MUDAP's interfaces for optimizing SLO fulfillment.

\subsection{Processing Service Execution}

\changed{To allow fine-granular management of allocated resources, each processing service is wrapped in an individual container.} Every second, a time-series DB scapes containers' resource utilization and service-specific metrics.
\changed{To address a service container, scaling agents provide the executing $host$ (or device), its service $type$, and container name. We summarize this through $s = \langle host,type,c\_name \rangle$. Thus, service containers can be accessed by local or remote scaling agents equally.}

\subsection{Elastic Service Adaptation}

By design, our multi-dimensional autoscaling platform can scale any parameter that can be dynamically adjusted during runtime\changed{, i.e., without container restart}. We broadly categorize parameters into: resource constraints and service configurations. Resource constraints limit the allocated resources per service container, such as the maximum scheduled CPU quota or RAM.
Service configurations adjust application-specific logic or functionality. For instance, a video detection service could adjust the size of a DNN model, or the size of the input tensor (i.e., the video resolution). Combined, these two sets form a service's \textit{elasticity parameters}, which can be adjusted through a REST API hosted within each container. Fig.~\ref{fig:MUDAP-architecture} shows the structure of such requests: a service endpoint, the elasticity parameter, and the parameter assignment. Thus, to change the input resolution of a video detection service, a potential API call could be \textit{/quality?resolution=1080}. This request is forwarded to the processing service; queries adjusting the CPU quota would go to the container. \changed{These changes do not require any container or application restart}.

To allow scaling agents interact with the elasticity parameters, services offer an API description; Table~\ref{tab:api-syntax} shows the respective syntax. For a specific service type, like "obj-detector", we define a list of elasticity strategies and their respective parameters. For each parameter, we specify the minimum and maximum value; assume a device could allocate up to 8 CPU cores (i.e., a CPU quota of $800.000.000$ns), the range would be $[1.0, 8.0]$. If the assignment exceeds the valid bounds, the value is clipped to the next valid assignment.

\begin{table}[t]
\caption{Syntax for defining multiple elasticity strategies per service; right column shown an example configuration}
\label{tab:api-syntax}
\begin{tabular}{lr}
\toprule
\textbf{Structural Entry} & \textbf{Example Value} \\
\midrule
\texttt{service type : list[str]} & \texttt{"obj-detector"} \\
\texttt{ > elasticity str. : list[str]} & \texttt{"resources"} \\
\texttt{$\;\;\;\;\;$ > url endpoint : str} & \texttt{"/resources"} \\
\texttt{$\;\;\;\;\;$ > query parameter : list[str]} & \texttt{"cores"} \\
\texttt{$\;\;\;\;\;\;\;\;\;$ > minimum value : float}  & \texttt{1.0} \\
\texttt{$\;\;\;\;\;\;\;\;\;$ > maximum value : float} & \texttt{8.0} \\
\bottomrule
\end{tabular}
\end{table}

At the later evaluation of MUDAP, we implement the different endpoints in the service logic and manually define the API description for the different services\footnote{The interested reader can find the full API description \href{https://github.com/borissedlak/elastic-workbench/blob/main/config/es_registry.json}{here}}. 

\section{Scaling Agent Design}
\label{sec:agent}

\begin{figure}
    \centering
    \includegraphics[width=0.9\columnwidth]{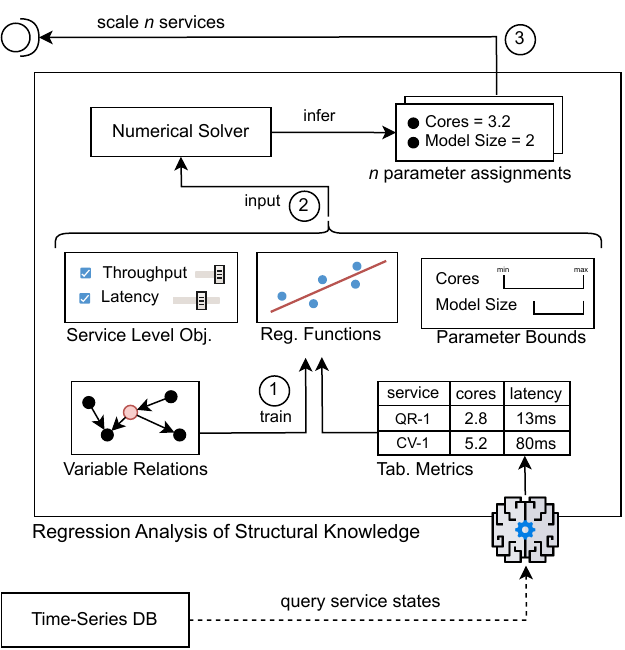}
    \caption{Conceptual sequence of RASK: \textcircled{\small 1} create a tabular structure from time-series data and train regression functions; \textcircled{\small 2} supply functions, SLOs, and parameter bounds to numerical solver; \textcircled{\small 3} optimize parameter assignments for all monitored services and autoscaled through MUDAP API.}
    \label{fig:rask-overview}
\end{figure}

While there exist various potential algorithms for implementing multi-dimensional scaling, we advocate model-based methods that allow interpreting the inferred scaling actions~\cite{sedlak_equilibrium_2024,chen_causeinfer_2019}. Further, given the dynamic and fast-paced nature of IoT streams, building scaling policies with few samples is desirable. 
%
To that extent, we present a tailor-made scaling agent that applies \underline{R}egression \underline{A}nalysis of \underline{S}tructural \underline{K}nowledge (RASK). RASK continuously improves its structural knowledge of the environment---expressed through continuous variable relations, which it uses to repeatedly optimize the SLO fulfillment through a numerical solver. 
As depicted in Fig.~\ref{fig:rask-overview}, RASK first \textcircled{\small 1} creates a tabular structure from the time-series data and trains the regression functions; it then \textcircled{\small 2} supplies these functions together with SLOs and parameter bounds to the numerical solver. Finally, \textcircled{\small 3} the solver produces parameter assignments for all monitored services, which are scaled through the MUDAP API. These steps---the \textit{RASK logic}---will be explained in the next subsections.

The RASK logic is executed in a continuous action-perception cycle---creating the \textit{RASK agent}. 
However, how often should this autoscaling cycle be executed? An argument for slower cycles would be the execution overhead, whereas faster cycles might help during highly dynamic request patterns~\cite{zhao_tiny_2022}. Also, changes to the processing environment are not immediately reflected, i.e., after an elasticity strategy, the system requires time to stabilize within the new configuration.
\changed{For example, the Kubernetes (k8s) Vertical Pod Autoscaler\footnotemark (VPA) by default recommends scaling actions every 1min; after applying an action, a 10min cooldown period is enforced. However, in our initial experiments, processing services (e.g., object detection) stabilized in less than 5s. This deserves dedicated optimization and analysis, but not in the scope of this paper. Hence, we simplify the decision and choose evaluation cycles of 10s---detached from any processing service.}
    
\footnotetext{\href{https://kubernetes.io/docs/tasks/configure-pod-container/resize-container-resources/}{Kubernetes VPA Description}, last accessed on \today}

\subsection{Observation of Processing Environment}

To collect the current state of the processing environment, the agent uses metrics from the time-series DB. 
Common time-series DBs (e.g., Prometheus) support metrics aggregation for stabilizing service states against natural fluctuations in the environment. 
While we set the evaluation interval to 10s, we also discussed that scaling actions take up to 5s for fully taking place; hence, we query a time series of the remaining 5s and consider the average. We repeatedly capture this information to create the training data ($D$) for the regression functions.
\changed{The agent's architecture thus favors fast paced adaptation cycles over a longer prediction horizon that uses more traces.}


\subsection{Regression Analysis of Structural Knowledge}


Building continuous variables relations allows interpolating between few samples captured through controlled interventions~\cite{sedlak_equilibrium_2024}. However, compared to conventional autoscalers~\cite{chen_survey_2018}, it also allows inferring continuous scaling actions---promoting fine-grained scaling decisions that converge towards a global optima~\cite{xing_fast_2022}.
In its core, RASK uses a simple numerical solver for finding an optimal parameter assignment. The novelty, however, lies in turning multiple elasticity parameters into convex functions, which can then be passed to the solver. This allows estimating the impact of different elasticity strategies and the respective SLO fulfillment. Hence, the crux for inferring satisfying assignments is developing an accurate but generalized regression model.

\subsubsection{Obtaining Structural Knowledge}

Creating the regression model and expanding it towards unseen parameter combinations faces an exploration-exploitation tradeoff: should the agent explore to improve its understanding of the environment (i.e., the accuracy of regression functions), or should it exploit known assignments to quickly rise SLO fulfillment, but potentially end up with a suboptimal configuration. To that extent, RASK supports two hyperparameters for tuning the exploration:
$\xi$---the length of an initial exploration phase, and  $\eta$---a percentage of Gaussian noise added to inferred parameter assignments, commonly used for RL-based exploration. Section~\ref{sec:evaluation} analyzes the effects of both parameters.

%
So how do we know which variables to relate? As there is rich literature on developing this knowledge through structural learning~\cite{kitson_survey_2023,vowels_ya_2021,sedlak_equilibrium_2024,sedlak_designing_2023_short}, we argue that existing work already addressed this research question. Hence, in the context of this paper, we supply directed variable relations---the structural knowledge (K)---according to expert knowledge. For a relation $k\in K$, which represents a variable's dependence on one or multiple other variables, we can extract the features ($X$) and target ($Y$) from the training data ($D$). As shown in Eq.~\ref{eq:regression}, we then fit a regression function ($\mathbf{w}^*$) to the data. According to the relation, the polynomial degree ($\delta$) can be customized---transforming the features $X$ into a higher order space. For example, $\delta=2$ produces a quadratic function.
\begin{equation}
\mathbf{w}^*(X,Y,\delta) := \arg\min_{\mathbf{w}} \sum_{i=1}^D \left( y_i - \mathbf{w}^\top \delta(x_i) \right)^2
\label{eq:regression}
\end{equation}


\subsubsection{Performing Regression Analysis}

\begin{algorithm}[t]
\caption{Regression Analysis of Structural Knowledge}
\label{alg:rask}
\begin{algorithmic}[1]

\REQUIRE services $(S)$, training data ($D$), Gauss. noise $(\eta)$, exploration $(\xi)$, polynomial degree $(\delta)$, knowl. structure ($K$), parameter bounds ($P$), constraints ($C$), SLOs ($Q$)
\ENSURE $A$ \COMMENT{assignments for elastic parameters}

\STATE $W \gets \varnothing$
\STATE $\text{rounds} \gets (\text{rounds} + 1)\; \textbf{if} \; \text{defined,} \; \textbf{else} \; 0$

\IF{$\text{rounds} < \xi$}
    \RETURN $\texttt{RAND\_PARAM}\;(P,c_{\text{max}})$
\ENDIF

\FOR{each $s \in S$ and $k \in K_s$}
\STATE $X; Y \gets D_{s}[k]$  \COMMENT{extract target features}
\STATE $W_s = \mathbf{w}^* (X, Y, \delta)$ \COMMENT{polynomial regression}
\ENDFOR

\STATE $A' \gets \texttt{SOLVE}\;(S,P,Q,W,C)$
\STATE $A \gets [\;\texttt{NOISE}(a', \eta) \mid a' \in A'\;]$

\RETURN $A$
\end{algorithmic}
\end{algorithm}


To optimize the service execution, the RASK agent adjusts the exposed elasticity parameters through a numerical solver. To that extent, we collect the parameter bounds ($P$) using the scaling platform's API description (see Table~\ref{tab:api-syntax}), a set of SLOs ($Q$) that define the desires QoE, and finally, the global resource constraints ($C$) governing the host. 
Using the numerical solver, the agent produces assignment ($A$) for each parameter ($p \in P$). For example, for a parameter $\textit{cores} \in P$ with bounds $[1.0, 8.0]$, a valid assignment $a \in A$ would be $\textit{cores} = 4.5$.

For a set of services ($S$), we now perform all steps of RASK as shown in Algo.~\ref{alg:rask}: the agent starts counting the number of autoscaling cycles (i.e., the \textit{rounds}) and keeps exploring as long as $\textit{rounds} < \xi$ stays true (Lines 2--4). Exploring, as shown by $\texttt{RAND\_PARAM}$ in Eq.~\eqref{eq:rand_ass}, means randomly assigning all parameters according to a uniform distribution, but within their bounds and global resource constraints. While this assumes zero prior knowledge, future work could benefit from more sophisticated exploration, e.g., avoid already visited states.
\begin{equation}
\begin{aligned}
\texttt{RAND\_PARAM} := & \;
\text{Draw } A \sim \mathcal{U}\left(p^{\min}, p^{\max}\right) \quad \\
\text{ s.t. } & \sum_{i=1}^{S}  p_i \leq C_p \\
& p^{\min} \leq p \leq p^{\max} \quad \forall p \in P_i
\end{aligned}
\label{eq:rand_ass}
\end{equation}
%

Once past the initial exploration period, the agent starts preparing the regression model ($W$)---a mere collection of all regression functions. For each service $s \in S$ and its structural relation $k \in K_s$, the agent does fit a regression function (Lines 6--9). In Line 10, the regression model $W$ is then passed to the numerical solver, together with the parameter bounds ($P$), the SLOs ($Q$), and the global constraints ($C$). As shown in Eq.~\eqref{eq:global_optimization}, the objective is to maximize the SLO fulfillment ($\phi$) for all SLOs ($q \in Q$) by assigning parameters within the bounds and global constraints. For SLOs concerning dependent variables, we estimate their value according to the regression model---$\mathbf{w}_i (p_i)$. Using gradient descent, a numerical solver (e.g., SLSQP~\cite{kraft_software_1988}) can optimize this objective.
%
\begin{equation}
\begin{aligned}
\texttt{SOLVE} :=
\max_{A} \quad & \sum_{i=1}^{S}\sum_{j=1}^{Q_i} \mathrm{\phi}(q_{j}, \; p_i \wedge \mathbf{w}_i(p_i))  \\
\text{s.t.} \quad & \sum_{i=1}^{S} p_i \leq C_p \\
& p^{\min} \leq p \leq p^{\max} \quad \forall p \in P_i
\end{aligned}
\label{eq:global_optimization}
\end{equation}

Finally, we apply Gaussian noise to each assignment. As shown in Eq.~\ref{eq:noise}, it first computes a standard deviation ($\sigma$) according to the assignment ($a'$) and the noise ratio ($\eta$); for example, $\text{cores}=4$ and $\eta=0.1$ produce $\sigma=0.4$. Each assignment is then shifted by its relative noise (or offset $o$).
\begin{equation}
\begin{aligned}
\texttt{NOISE}\;(a,\eta) := a + o; \; o \sim \mathcal{N}(0, \sigma) \text{ and } \sigma = (a \times \eta)^2
\end{aligned}
\label{eq:noise}
\end{equation}
\noindent
After reaching Line 12, the noisy parameter assignment is returned. Otherwise, if the agent was still exploring, Line 5 already returned the randomized assignment. 

\vspace{5pt}
\subsubsection{Tuning the Numerical Solver}
\label{subsubsec:solver-tuning}

A common choice is to start a numerical solver from a randomized, or averaged parameter assignment. Using this strategy, our initial experiments showed outliers in the duration of the solver.
To minimize the risk of exceeding the autoscaling interval, we implement a caching mechanism: instead of starting the numerical solver from a random or default state, we cache the last parameter assignment to kickstart the solver. 
However, we hypothesize that this might trap the solver in a local optima, hence, we analyze the implications of the caching in Section~\ref{subsec:evaluation-e1}.

\subsection{Service Autoscaling}

Given the assignment ($A$) provided by RASK, the agent now has to scale the services. For this, it uses the API of the autoscaling platform to adjust the services according to the parameter assignments. For each assignment $a \in A$, this means sending a request to the exposed REST API; the structure of these requests was explained alongside Fig.~\ref{fig:MUDAP-architecture}. 

This concludes the autoscaling cycle of the RASK agent, which is continuously executed every 10 seconds---the evaluation interval. However, as the upcoming evaluation also shows, the agent usually finished the autoscaling in a fraction of that time. Hence, it minimizes the resource overhead by remaining idle until invoked again.

\section{Evaluation}
\label{sec:evaluation}

To show the full potential of our approach, we evaluate the autoscaling platform and RASK empirically. First, we describe the implementations of our prototype and the three processing services embedded. Afterward, we describe the conducted experiments, which analyze the training time of RASK, its performance in comparison with SOTA autoscalers, and its scalability for larger problem sizes. We provide our implementations and experimental results on GitHub\footnotemark.

\footnotetext{\href{https://github.com/borissedlak/elastic-workbench/tree/tsc-submission}{Github repository} with experimental results in \href{https://github.com/borissedlak/elastic-workbench/tree/tsc-submission/experiments/tsc}{subfolder}}

\subsection{Prototype Implementation}

We implement MUDAP and RASK both in Python--- building heavily on existing tools and packages. For MUDAP, we containerize applications in Docker, which allows dynamically scaling the container limits (e.g., CPU quota) through the Docker API. The REST API itself consists of a simple HTTP server executed within the container; requests to the REST API are either executed over the Docker API, or routed to the respective service. Lastly, we use Prometheus\footnotemark as our time-series DB, which supports a wide range of queries and metrics aggregation.
For RASK, we train regression functions between dependent variables using sklearn~\cite{pedregosa_scikit-learn_2018_short} with a polynomial degree $\delta=2$. We use scipy~\cite{virtanen_scipy_2020_short} as our numerical solver, which internally uses SLSQP~\cite{kraft_software_1988}---an algorithm that supports bounded parameters and global resource constraints. \changed{To capture RASK's overhead, we also wrap it in a container.}

\subsection{Service Implementation}

\begin{figure}[!t]
\vspace{-10pt}
    \subfloat[QR Detector (\textit{QR})]{%
    \includegraphics[width=0.318\columnwidth]{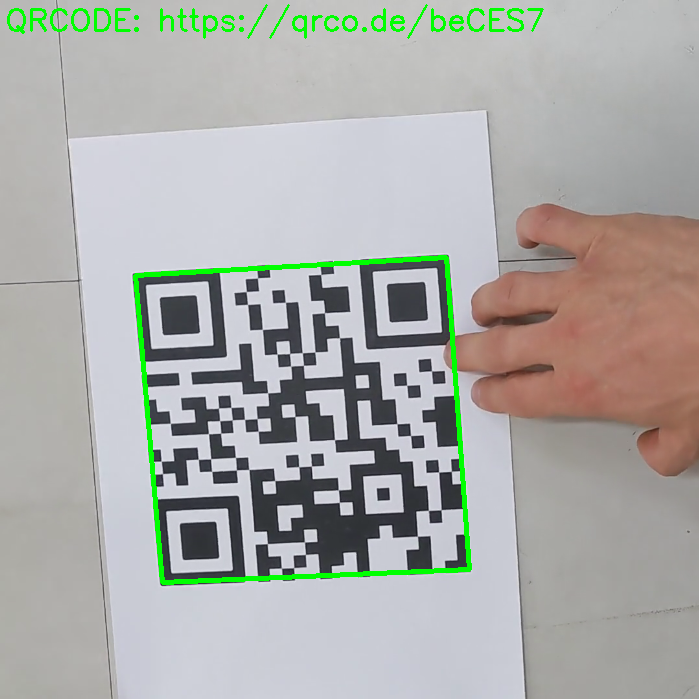}%
    \label{fig:service-demo-1}%
    }%
    \hspace{0.0135\columnwidth}
    \subfloat[Object Detector (\textit{CV})]{%
        \includegraphics[width=0.318\columnwidth]{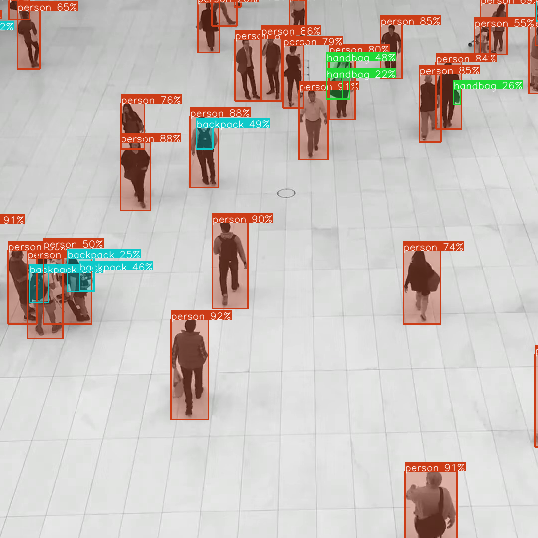}%
    \label{fig:service-demo-2}%
    }
    \hspace{0.0135\columnwidth}
    \subfloat[Lidar Renderer (\textit{PC})]{%
        \includegraphics[width=0.318\columnwidth]{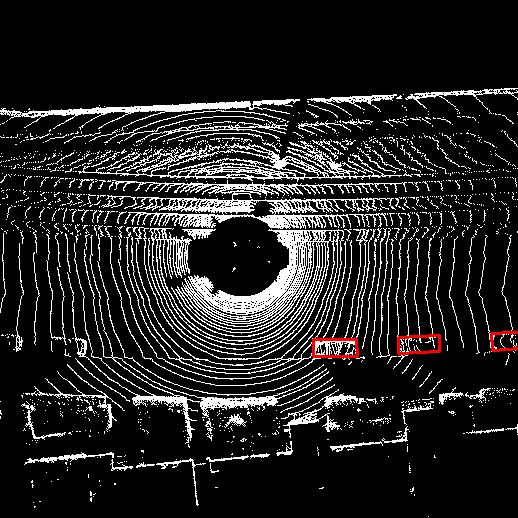}%
    \label{fig:service-demo-3}%
    }
    \caption{Output of the three implemented processing services.}
    \label{fig:service-demos}
\end{figure}

We develop three processing services that will be executed within the MUDAP platform.
Fig.~\ref{fig:service-demos} shows the demo output for each of the three services, namely, we implement: (a) a QR code reader that detects QR code within video frames through OpenCV~\cite{opencv_opencv_2024}; (b) a CV service that uses Yolov10~\cite{wang_yolov10_2024} for detecting and classifying objects in video frames; and (c) a point cloud renderer based on the Kitti dataset~\cite{geiger_are_2012} that draws a mobile map around a moving vehicle.

\changed{All three services operate in cycles of 1000 ms---every second, they retrieve data items from the buffer (i.e., either video frames or point cloud binaries), and process as many items as possible.} Upon completion, or after exceeding 1000 ms, the number of processed items (i.e., the \textit{throughput}) and other metrics are scraped by the time-series DB. 

\footnotetext{\href{https://prometheus.io/}{Prometheus Time-Series DB}, last accessed on \today}

\paragraph{Elasticity Parameters}

For ensuring SLOs, each service supports its own set and range of elasticity parameters. We summarize this in Table~\ref{tab:slo-thresholds}; for example, the QR service features three variables: \textit{cores}, \textit{data quality}, and \textit{completion} rate. While \textit{cores} and \textit{data quality} can be adjusted as elastic parameters, \textit{completion} is calculated as
\begin{equation}
    \textit{completion} = \textit{throughput}\; /\; RPS
\label{eq:completion-rate}
\end{equation}

While all three services can adjust their \textit{data quality}, the parameter refers to different aspects and different ranges: for QR and CV, it represents the size of the video frame, while for PC it defines the maximum range of processed points.
\changed{Notably, rescaling \textit{data quality} causes no additional delay.}
The CV service exposes one more parameter---\textit{model size}---which allows switching from YOLOv10\underline{n} up to v8\underline{l}, corresponding to the range $[1,4]$. \changed{This change is performed at runtime by loading the new model into memory and swapping the ONNX model execution.} To support variable input size (i.e., \textit{data quality}), we export Yolov10 with the \textit{dynamic} preset. Still, the size of input tensors must be a multiple of 32---otherwise, the service API clips it to the next valid assignment.

\begin{table}[t]
\setlength{\tabcolsep}{4.5pt}
\centering
\caption{Service variables for the QR, CV, and PC service; SLO targets and their \textbf{w}eights (i.e., importance) included. }
\begin{tabular}{cllc|lc|c}
\toprule
\textbf{Service} & \textbf{Variable} & \textbf{Descript.} & \textbf{Range} & \textbf{SLO} & \textit{w} & \textbf{Step}\\
\midrule
\multirow{3}{*}{QR} 
    & \textit{cores} & CPU quota& $(0,8)$ & -- & -- & float\\
    & \textit{data quality} & Image size & $[10^2,10^3]$ & $\geq 800$ & 0.5 & $\pm 1$\\
    & \textit{completion} & Rate finish & $[0,1]$ & $\geq 1.0$ & 1.0 & --\\
\midrule
\multirow{4}{*}{CV} 
    & \textit{cores} & CPU quota & $(0,8)$ & -- & -- & float\\
    & \textit{data quality} & Image size & $[128,320]$ & $\geq 288$ & 0.2 & $\pm 32$\\
    & \textit{model size} & Yolov10[n/.]& $[1,4]$ & $\geq 3$ & 0.2 & $\pm 1$\\
    & \textit{completion} & Rate finish & $[0,1]$ & $\geq 1$ & 1.0 & --\\
\midrule
\multirow{3}{*}{PC} 
    & \textit{cores} & CPU quota & $(0,8)$ & -- & -- & float\\
    & \textit{data quality} & Lidar range & $[6,60]$ & $\geq 40$ & 0.5 & $\pm 1$\\
    & \textit{completion} & Rate finish & $[0,1]$ & $\geq 1$ & 1.0 & --\\
\bottomrule
\label{tab:slo-thresholds}
\end{tabular}
\end{table}

As discussed before, the scaling agent builds regression functions between dependent variables---expressed through structural knowledge (K). Given the variables in Table~\ref{tab:slo-thresholds}, we define $K$ for the three services as 
\begin{equation}
\begin{aligned}
    K_{QR}, K_{PC} = \{cores,data\;quality\} \rightarrow tp_{\text{max}}\\
    K_{CV} = \{cores,data, model\;quality\} \rightarrow tp_{\text{max}}
    \label{eq:K}
\end{aligned}
\end{equation}
which expresses the maximum expected throughput ($tp_{\text{max}}$) according to the service configuration. The $tp_{\text{max}}$ is calculated independently of the current RPS, solely using the processing \textit{latency}. When building the regression model, this allows extrapolating from known configurations to potentially higher RPS; thus answering the question: \textit{could I have processed more data with the current configuration}? Still, the solver uses the $tp_{\text{max}}$ equally for calculating the \textit{completion} SLO.

\vspace{5pt}
\paragraph{Service Level Objectives}
To ensure QoE, all services aim for high \textit{data quality}. In particular, this means high video resolutions for the QR service (i.e., $\geq 800\text{px}$) and the CV service (i.e., $\geq 288\text{px}$). Apart from that, the CV service aims for high detection accuracy through a large model size (i.e., $\geq$ v8\underline{m}). 
\changed{Further, all services aim to fulfill 100\% completion rate, as known from Eq.~\eqref{eq:completion-rate}---this means, that services must process all items stored in the buffer before frames can time out after 1000ms.}
While there is no SLO target on the allocated \textit{cores}, they impact the service \textit{throughput} (cfr Eq.~\ref{eq:K}). Hence, the goal is learning a globally-optimal resource division.

\changed{A resource-constrained Edge device will fail to reach the desired \textit{throughput} under increasing RPS, particularly when splitting resources among services. To that extent, our scaling agent can trade off qualitative aspects of the application (e.g., \textit{data quality} or \textit{model size}) to sustain a minimum \textit{throughput}---this behavior can be configured through the associated SLO weights. By assigning the highest weight to the \textit{completion} SLO (i.e., $w=1.0$), we create a hierarchy of qualitative aspects that must be ensured by all means, and such that can be traded off during periods of high load. Additionally, these weights can be tuned to shift importance between services---needed to orchestrate between multiple concurrent tenants.}

\vspace{5pt}
\paragraph{Default Parameter Assignments and Request Load}

At the beginning of experiments or an experimental run, we reset the processing services (i.e., all their elasticity parameters) to the default states shown in Table~\ref{tab:service-defaults}. Initially, each service is allocated a equal share from the device resources: $C\; /\; |S|$. The other elastic parameters are assigned with the half range of their bounds $\Rightarrow(p^{\max} - p^{\min}) \; / \; 2$. Lastly, we also set the incoming RPS to a default value; this assignment is adjusted at later experiments according to the service load.

\begin{table}[h]
\centering
\caption{Default RPS and elastic parameter assignments; between experimental runs, we reset services to these values.}
\begin{tabular}{cclr}
\toprule
\textbf{Service} & \textbf{Default RPS} & \textbf{Variable} & \textbf{Value} \\
\midrule
\multirow{2}{*}{QR} 
    & \multirow{2}{*}{80} 
    & \textit{cores} & $2.6$ \\
    & & \textit{data quality} & 550 \\
\midrule
\multirow{3}{*}{CV} 
    & \multirow{3}{*}{5} 
    & \textit{cores} & $2.6$ \\
    & & \textit{data quality} & 224 \\
    & & \textit{model size} & 3 \\
\midrule
\multirow{2}{*}{PC} 
    & \multirow{2}{*}{50} 
    & \textit{cores} & $2.6$ \\
    & & \textit{data quality} & 30 \\
\bottomrule
\label{tab:service-defaults}
\end{tabular}
\end{table}

\subsection{Experiment Setup \& Results}

To evaluate the performance and viability of our approach from various angles, we design the following six experiments: \textbf{E1} analyzes the convergence of RASK under different hyperparameter settings; \textbf{E2} analyzes the effect of different polynomial degrees on the regression functions; \textbf{E3} compares the performance of RASK with SOTA baselines. We then analyze the scalability of RASK for (\textbf{E4}) increasing numbers of elasticity strategies, (\textbf{E5}) the effect of caching parameter assignments, and (\textbf{E6}) increasing processing services.
For each experiment, we outline the setup of the processing environment, and then provide the respective results.

All experiments were conducted on a VM with an AMD EPYC 7742 CPU and 12 GB of RAM. We prefer this setup over a physical Edge device (e.g., NVIDIA Jetson\footnote{\href{https://www.nvidia.com/en-gb/autonomous-machines/embedded-systems/}{NVIDIA Jetson Documentation}, last accessed on \today}), because this allows changing the resource constraints between experiments---needed for \textbf{E6}. Although RASK is designed to optimize any parameterizable hardware allocation, we focus our evaluation only on allocating 8 CPU $cores$. To the present day, the GPU cannot be limited for individual Docker containers\footnotemark. Also, we found that all our service configurations had a combined RAM usage of less than 4GB, which would not exhaust an Edge devices like NVIDIA Jetson. Hence, integrating other hardware dimensions remains for future work.

\footnotetext{\href{https://docs.docker.com/engine/containers/resource_constraints/}{Docker Limits} without GPU partitioning, last accessed \today}

\vspace{5pt}
\subsubsection{Training Duration of RASK (\textbf{E1})}
\label{subsec:evaluation-e1}

The exploration in RASK can be tuned through two hyperparameters: the number of rounds ($\xi$) in which the agent explores randomly, and the amount of Gaussian noise ($\eta$) added to actions. In RL, $\eta = 0.1$ is a common values, whereas we estimate $\xi$ according to previous work~\cite{sedlak_equilibrium_2024}. Hence, we define three assignments of $\xi = \{0,10,20\}$ and two of $\eta = \{0,0.1\}$, and use their product for evaluating 6 different hyperparameter combination. Each configuration is executed for 10min and the agent operates in cycles of 10s; hence, 60 iterations with the environment. To stabilize our findings, we execute 5 repetitions per config.

\paragraph{Results}

\begin{figure}[t]
    \centering
    \includegraphics[width=1.01\columnwidth]{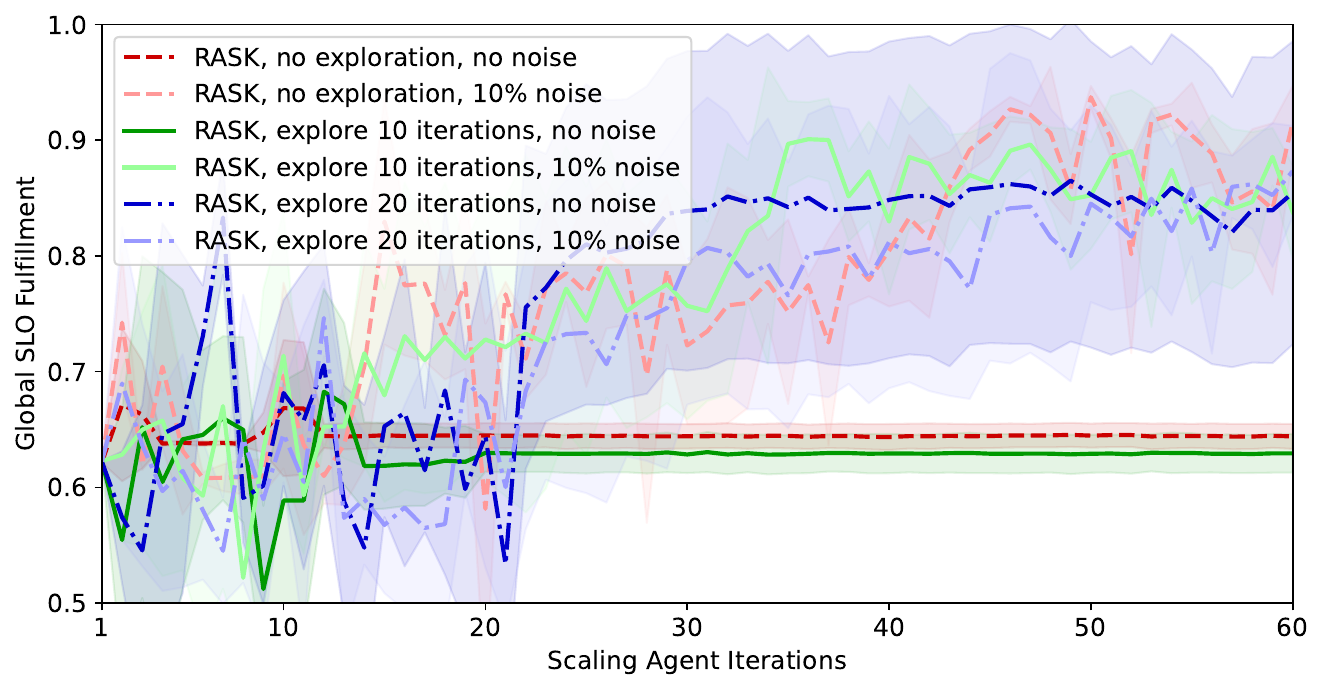}
    \caption{Training RASK scaling agent in 60 iterations (=10min); different runs show how training convergence and SLO fulfillment are impacted by an extended initial exploration phase, and Gaussian noise added to the inferred scaling actions}
    \label{fig:E1_SLO_F}
\end{figure}

For each combination of hyperparameters, Fig.~\ref{fig:E1_SLO_F} shows the globally-weighted SLO fulfillment averaged over all three services, which we calculate according to
\begin{equation}
    (\sum_{i=1}^{S} (\sum_{j=1}^{Q_i} \phi_j \times w_j) \;/ \sum_{j=1}^{Q_i} w_j) \;/\; |S|
\end{equation}

We observe three things: (1) without a dedicated exploration phase $\{\xi =0\}$, or merely a short one $\{\xi =10\}$, the agent did not find satisfying parameter assignments. Nevertheless, (2) all noisy configurations with $\{\eta =0.1\}$ led to satisfying assignments; however, their high fluctuation, even in late iterations, indicates that the noise should decay as the performance converges. Finally, (3) exploring merely for 20 iterations appears sufficient for developing a stable regression model; however, the high standard deviation also indicates that in one run it could not find a highly-satisfying assignment. 

Notice, that all remaining experiments will use the metrics of \textbf{E1} $\{\xi =20, \eta=0\}$ for pretraining the RASK agent.

\subsubsection{Polynomial Degree of RASK (\textbf{E2})}


While we started our evaluations using a default polynomial degree of $\delta = 2$, we hypothesized that a service-specific degree might improve the generalization to unseen data. To that extent, we analyze the impact of different polynomial degrees on the mean squared error (MSE) of a 20\% test split; \changed{for this, we use exclusively training data from \textbf{E1}}. This analysis also mitigates the risk of overfitting when using high polynomial degrees.


\paragraph{Results}

For each service, Table~\ref{tab:degree-mse} shows the MSE for using polynomial degrees from 1 to 6: QR and PC are best expressed through a polynomial function of $\delta=4$, while CV best uses a linear function (i.e., $\delta=1$). In particular, the CV service decreased its MSE by a factor of \textbf{2.4} compared to the default degree. Since our expert knowledge of the structure $K$ did not include the optimal function degree, we see strong evidence for using service-specific degrees in future work.

\begin{figure}[t]
\vspace{-6pt}
    \subfloat[QR Detector (\textit{QR})]{%
    \includegraphics[width=0.318\columnwidth]{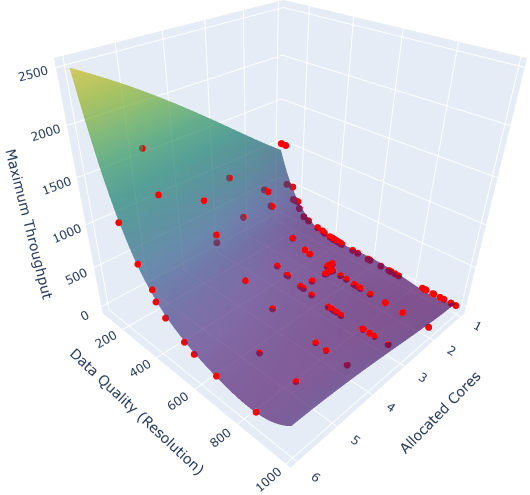}%
    \label{fig:regression-qr}%
    }%
    \hspace{0.0135\columnwidth}
    \subfloat[Object Detector (\textit{CV})]{%
        \includegraphics[width=0.318\columnwidth]{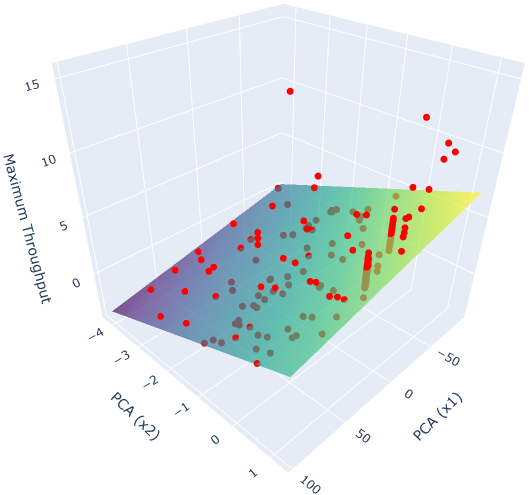}%
    \label{fig:regression-cv}%
    }
    \hspace{0.0135\columnwidth}
    \subfloat[Lidar Renderer (\textit{PC})]{%
        \includegraphics[width=0.318\columnwidth]{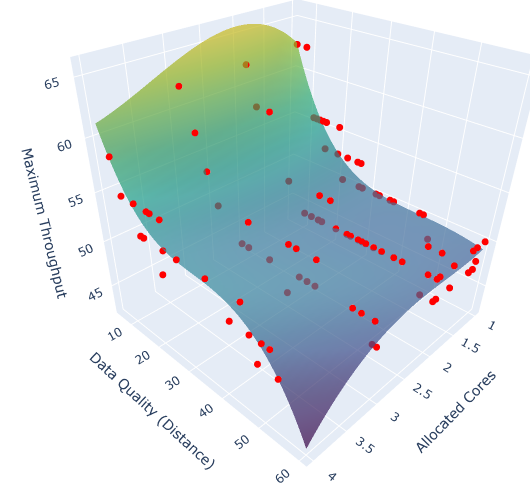}%
    \label{fig:regression-pc}%
    }
    \caption{Structural knowledge of the expected service \textit{throughput} under different service configurations; the degree ($\delta$) of the polynomial function is determined according to the least MSE from Table~\ref{tab:degree-mse}: QR and PC use $\delta = 4$, and CV $\delta=1$.
    }
    \label{fig:regression-functions}
\end{figure}

Further, we visualizing the regression models using each service's optimal degree. 
\changed{In Fig.~\ref{fig:regression-functions} we show the expected \textit{throughput} (i.e., the $tp_{\text{max}}$) for different parameters assignments. The RASK agent will consider these dynamics to build a tradeoff between parameters.}
While QR and PC only feature two parameters---making them easily displayed in 3D---the CV also allows adjusting the Yolov10 \textit{model size}, so we reduce its features through PGA. We note that the service \textit{throughput} is always highly impacted by \textit{data quality} and \textit{cores}, except for the PC service, which indicates poor parallelization.

\begin{table}[t]
\centering
\caption{\changed{Model accuracy} for fitting a regression function to the training data created in \textbf{E1}. Cells show the Mean Squared Error (MSE) on a 20\% test split; green (bright) cells indicate the optimal polynomial degree for fitting the function, and orange (dark) cells the default setting in our experiments.} 
\label{tab:degree-mse}
\begin{tabular}{cccc}
\toprule
\textbf{Poly. Degree} & \textbf{QR Detector} & \textbf{CV Analyzer} & \textbf{PC Visualizer} \\
\midrule
1 & 19114.41 & \cellcolor{green!14}3.65 & 3.85 \\
\cellcolor{orange!38}2 & \cellcolor{orange!38}6315.82  & \cellcolor{orange!38}4.22 & \cellcolor{orange!38}2.53 \\
3 & 4290.81  & 4.64 & 2.27 \\
4 & \cellcolor{green!14}2650.09  & 4.25 & \cellcolor{green!14} 2.17 \\
5 & 4168.13  & 4.98 & 2.23 \\
6 & 5740.43  & 4.66 & 2.28 \\
\bottomrule
\end{tabular}
\end{table}

\vspace{5pt}
\subsubsection{\changed{Performance Comparison for Dynamic Workloads (\textbf{E3})}}
\label{subsubsec:E3}


To analyze the performance of RASK further, we compare it to two baseline agents. \changed{During the experiments, agents will be evaluated according to their capability to ensure SLO fulfillment during two dynamic and unseen request pattern.}

\paragraph{Baselines}

Common autoscalers often use the default k8s implementation, or a custom RL algorithm~\cite{quattrocchi_autoscaling_2024_short,gari_reinforcement_2021}. To that extent, we compare the performance of RASK against the following two baselines:

\begin{enumerate}

    \item[\textbf{VPA}] -- Replicates the behavior of the k8s VPA. For each service container, it aims to maintain a resource slack of 5\% to 15\%~\cite{zhao_tiny_2022}; hence, it consumes between 85\% and 95\% of the maximum scheduled CPU quota. If the service execution violates these bounds, the \textbf{VPA} agent adjusts the allocated $\textit{cores} \pm 0.25$. If all available resources are allocated, they can only be reassigned once released.

    \vspace{5pt}
    
    \item[\textbf{DQN}] -- \changed{Approximates Q-values for discrete state-action pairs. To support service-specific scaling policies, services are modeled through separate Deep Q-Networks (DQNs)~\cite{xue2022meta, xu_coscal_2022,mayerhofer2025hpaqt}.} Models are pre-trained jointly within a shared Gymnasium\footnotemark environment: given an action, the environment estimates the expected state and reward (i.e., SLO fulfillment); for this, it reuses RASK's regression model. The \textbf{DQN} agent has access to all available elasticity dimensions; however, to simplify the action space, it infers a single action per service.
    \vspace{3pt}

    \footnotetext{\href{https://gymnasium.farama.org/api/env/}{Gymnasium Env} for pretraining RL agents, last accessed \today}
    
\end{enumerate}

\changed{The DQN and the RASK agent are both pretrained on one static request pattern---the default RPS in Table~\ref{tab:service-defaults}; for this, the agents can only use the initial observations from \textbf{E1}.}
While the baseline agents scale services according to their own internal logic, they likewise operate on the MUDAP platform: all agents query service states from the time-series DB and adjust elasticity parameters through the REST API.
Also, all agents equally start the service execution from the default values in Table~\ref{tab:service-defaults}. Given that the VPA agent cannot adjust \textit{data quality}, this makes it technically impossible to meet the associated SLO and maximize fulfillment; however, our experiments showed that the \textit{CV} service would otherwise not reach a $\textit{throughput} \geq 1\text{ fps}$, and hence perform even worse.

\paragraph{\changed{Dataset}}

\begin{figure}[t]
\vspace{-8pt}
    \centering
    \subfloat[Bursty]{%
        \includegraphics[width=0.499\columnwidth]{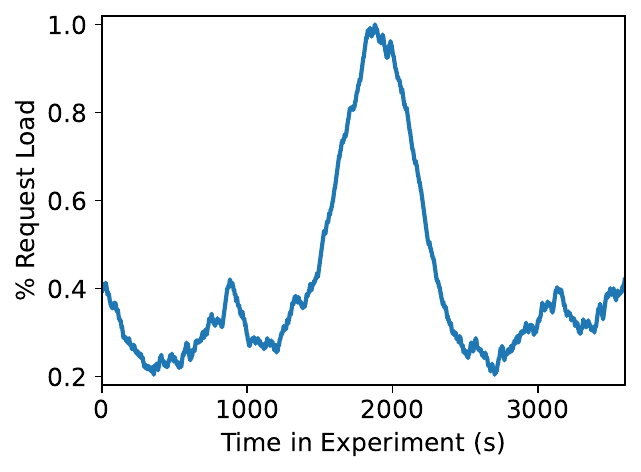}%
        \label{subfig:pattern-bursty}%
    }%
    \subfloat[Diurnal]{%
        \includegraphics[width=0.499\columnwidth]{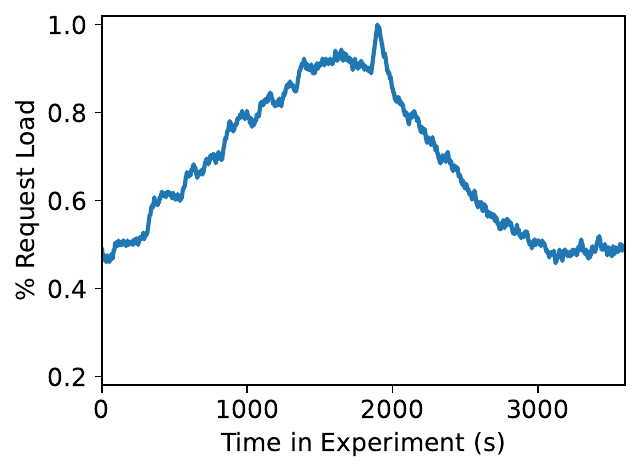}%
        \label{subfig:pattern-diurnal}%
    }%

    \caption{Two common request patterns in Google Cluster production environments~\cite{wilkes_google_2020,wang_autothrottle_2024}. We scale the relative request load to a maximum of \textbf{100} RPS for the \textit{QR} service and \textbf{10} RPS for the \textit{CV} service. Each agent is evaluated under both load patterns in separate experiments, each lasting 1 hour.}
    \label{fig:req-pattern}

    \vspace{6pt}

    \subfloat[Bursty]{%
        \includegraphics[width=0.499\columnwidth]{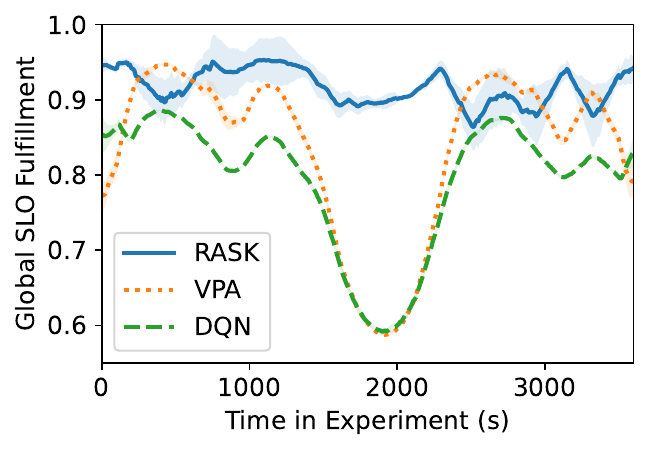}%
        \label{subfig:E2_SLO_F-bursty}%
    }%
    \subfloat[Diurnal]{%
        \includegraphics[width=0.499\columnwidth]{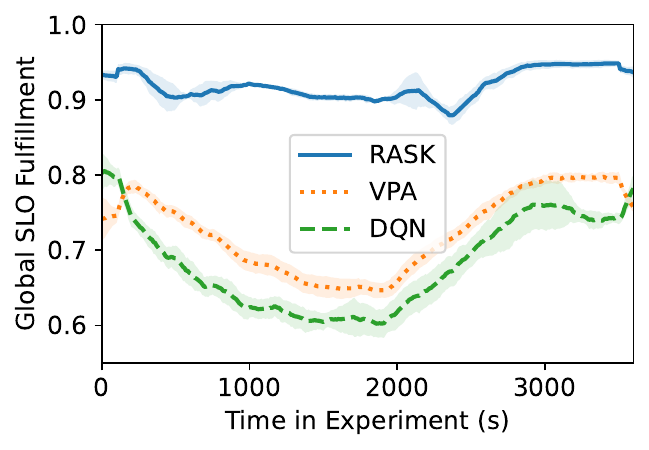}%
        \label{subfig:E2_SLO_F-diurnal}%
    }%

    \caption{Global SLO fulfillment across all three services during the \textit{Bursty} and \textit{Diurnal} request patterns. The resulting performance closely aligns with the peaks in the request load, which our method (i.e., RASK) sustains with the fewest violations.}
    \label{fig:E2_SLO_F}
\end{figure}

\changed{We evaluate the agents' performance using two common request patterns from the Google Cluster production environments~\cite{wilkes_google_2020,wang_autothrottle_2024}. These traces exhibit dynamically changing user behavior and unprecedented bursts---forming a drastic change compared to the static, pretrained pattern.} Fig.~\ref{fig:req-pattern} visualizes both request pattern---each capturing a duration of one hour. Notice, how the relative request load is scaled to a maximum of 100 RPS for the \textit{QR} service and 10 RPS for our \textit{CV} service. Hence, it becomes increasingly more difficult to fulfill the SLOs from Table~\ref{tab:slo-thresholds}. For the PC service, however, we assume a constant RPS---reasonable for an individual vehicle client. We run 5 repetitions for each evaluated agent.

\paragraph{Results}

Fig.~\ref{fig:E2_SLO_F} shows the mean globally-weighted SLO fulfillment for the request pattern, including the standard deviation---we added the detailed SLO fulfillment and resource allocation as Appendices. We observe: (1) in periods of high load (i.e., $\text{load} \geq 0.4$) there is a particularly large gap between RASK and the baselines; while (2) in periods of lower load (i.e., $\text{load} < 0.4$) the agents reported comparable SLO fulfillment, with VPA even outperforming RASK at $\text{load} = 0.2$. While the VPA's low SLO fulfillment supports our push towards multiple elasticity strategies, the DQN's low fulfillment requires additional analysis. To date, we attribute it to the fact that the DQN was not trained for different RPS, but neither was the RASK agent in \textbf{E1}. Hence, we conclude that RASK generalizes better under these circumstances.

\vspace{5pt}
\subsubsection{Complexity of RASK -- Elasticity Challenges (\textbf{E4})}

Our final three experiments assess the complexity and scalability of our approach for increasingly larger problem sizes. 
%
%
We compare three instances of our RASK agent: operating in 1 dimension---adjusting only \textit{cores}; in 2 dimensions---adjusting \textit{cores} and \textit{data quality}; and in 3 dimensions---adding \textit{model size} for the CV services. \changed{We evaluate these agents for the \textit{Diurnal} request pattern---5 repetitions each---and capture RASK's overhead as in \cite{usman_lightweight_2025}: CPU, RAM, and performed GFLOPS\footnotemark.}

\footnotetext{To count GFLOPS per scaling cycle, we capture the RASK container's $\texttt{fp\_ret\_sse\_avx\_ops.all}$ and divide by the \# of performed cycles.}

\paragraph{Results}

\begin{figure}[t]
    \subfloat[Runtime Complexity]{%
    \includegraphics[width=0.499\columnwidth]{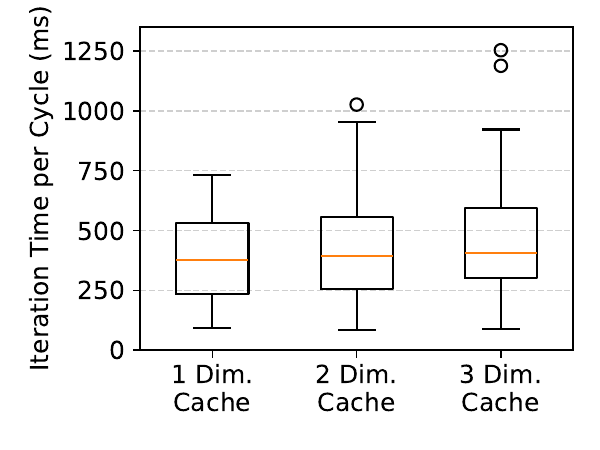}%
    \label{subfig:E3_1_complexity_Cache}%
    }%
    \subfloat[SLO Fulfillment]{%
        \includegraphics[width=0.499\columnwidth]{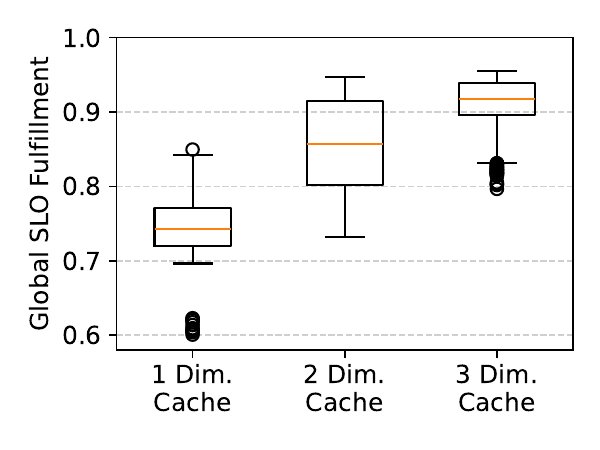}%
    \label{subfig:E3_1_SLO_F_Cache}%
    }
    \caption{Runtime complexity and SLO fulfillment of RASK with increasing numbers of elasticity strategies (i.e., \textbf{Dim}ensions). Agents are evaluated on the \textit{Diurnal} request pattern.}
    \label{fig:E3_1_Performance_Cache}
\end{figure}

\changed{Fig.~\ref{fig:E3_1_Performance_Cache} shows the distributions of the RASK agent's runtime (ms) for inferring a parameter assignment, together with the SLO fulfillment throughout the runs.} Considering these plots, we observe that increasing the number of elasticity strategies improved the global SLO fulfillment from a median value of $\mathbf{0.75}$ for 1 dimension up to $\mathbf{0.92}$ for 3 dimensions. Further, regarding the agent's runtime, we report a minor increase between a median of $\mathbf{357ms}$ for 1 dimension compared to $\mathbf{395ms}$ for 3 dimensions. In comparison, baseline agents required on average less than $\mathbf{50ms}$.

\changed{To analyze the runtime complexity of RASK's individual steps (i.e., training, inference, and orchestration) we create an average latency breakdown in Table~\ref{tab:latency-breakdown}: for 3 dimensions, the agent's latency is dominated by the training part ($\textbf{90\%}$). As a comparison, we analyze the latency during the exploration phase in \textbf{E1}, where the agent does not yet train regression models, but infers random parameter assignments.}

Finally, Table~\ref{tab:rask-overhead}, presents the resource overhead introduced by the the RASK agent: while the permanently occupied RAM stays constant across increasing elasticity dimensions, the CPU usage and performed FLOPS scale with the dimension.



%
\begin{table}[t]
\centering
\caption{\changed{Latency breakdown for RASK at runtime: during regular operation model training dominates the overall latency; during exploration only random assignments are inferred.}} 
\label{tab:latency-breakdown}
\begin{tabular}{r|rrr|r}
\toprule
\textbf{Phase} & \textbf{Train model} & \textbf{Infer params.} & \textbf{Orchestrate} & \textbf{Total} \\
\midrule
Explore & --- & $<1$ ms (2 \%)  & 10 ms (98\%) & 10 ms \\
Exploit & 355 ms (90 \%) & 30 ms (7 \%) & 10 ms (3 \%) & 395 ms \\
\bottomrule
\end{tabular}
\end{table}
\changed{
\begin{table}
    \caption{\changed{Runtime overhead of RASK autoscaler: higher elasticity dimensions mildly increase the CPU demand and the average GFLOPS performed in every scaling cycle.}}
    \label{tab:rask-overhead}
    \centering
    \begin{tabular}{r|cc|cc}
    \toprule
        \multirow{2}{*}{Setup} & \multicolumn{2}{c|}{Occupied Permanently} & \multicolumn{2}{c}{Average per Cycle} \\
        & CPU & RAM & GFLOPS\\
    \midrule
        1 Dim. & 0.024 cores & 105 MB & 0.062 GFLOPS \\
        2 Dim. & 0.033 cores & 105 MB & 0.090 GFLOPS \\
        3 Dim. & 0.043 cores & 107 MB & 0.135 GFLOPS \\
    \bottomrule
    \end{tabular}
\end{table}}

\vspace{5pt}
\subsubsection{Caching of RASK Assignments (\textbf{E5})}

In Section~\ref{subsubsec:solver-tuning} we stated our concerns about how caching parameter assignments between RASK iterations might trap the numerical solver in a local optimum. To that extent, we use the results of \textbf{E4}---the three RASK agents with increasingly more elasticity strategies---and compare it with three agents that have caching disabled.

\paragraph{Results}

\begin{figure}[t]
    \subfloat[Runtime Complexity]{%
    \includegraphics[width=0.499\columnwidth]{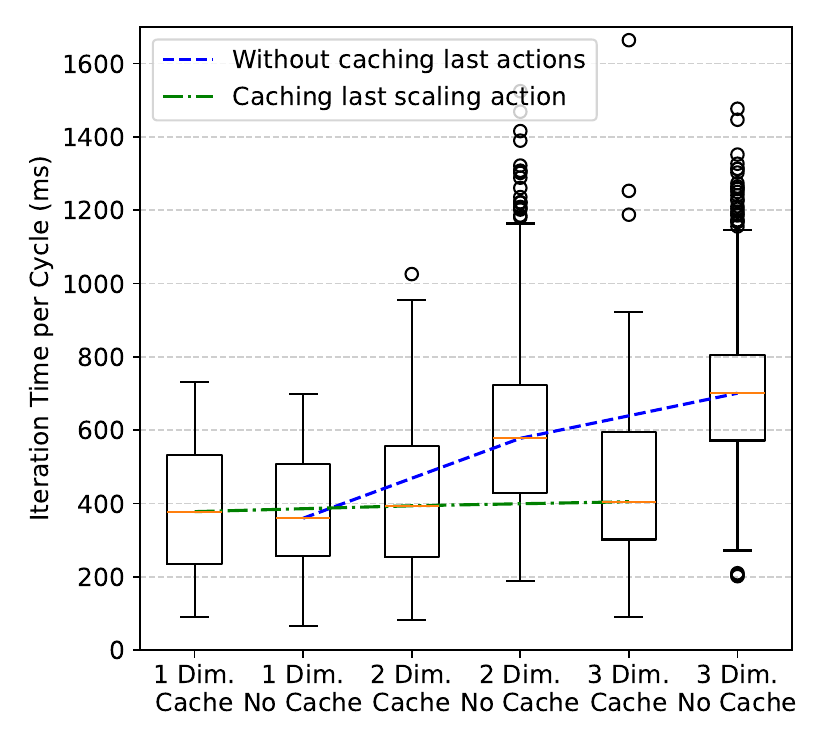}%
    \label{subfig:E3_1_complexity}%
    }%
    \subfloat[SLO Fulfillment]{%
        \includegraphics[width=0.499\columnwidth]{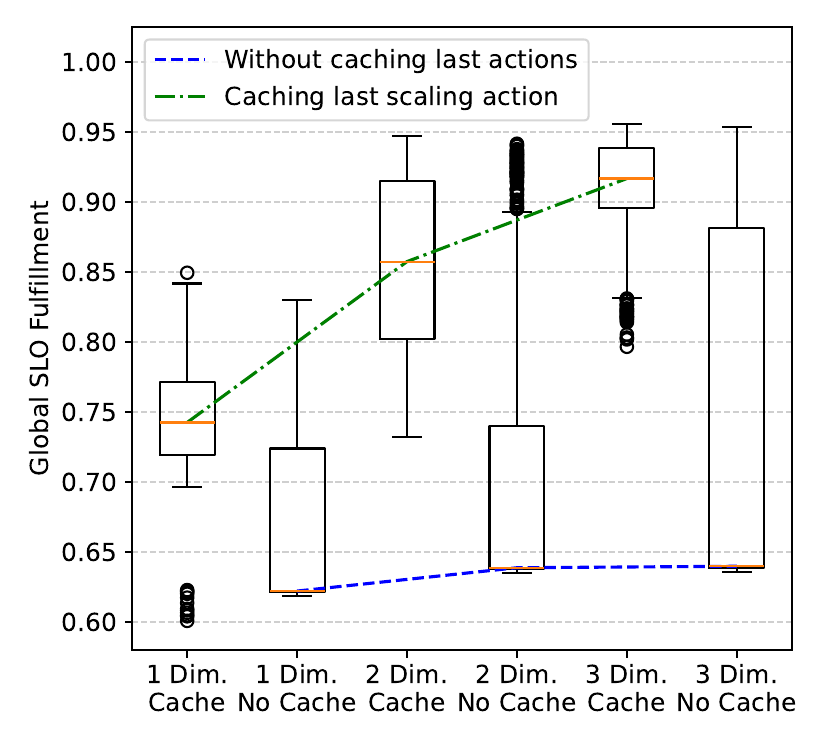}%
    \label{subfig:E3_1_SLO_F}%
    }
    \caption{Runtime complexity and SLO fulfillment of RASK with / without caching last parameter assignment. Agent was evaluated on \textit{Diurnal} pattern with different \textbf{Dim}ensions.}
    \label{fig:E3_1_Performance}
\end{figure}

Fig.~\ref{fig:E3_1_Performance} shows the runtime and SLO fulfillment for RASK agents with caching enabled or disabled. Considering these plots, we reason that (1) caching the last parameter assignment ensured stable runtime $\leq\mathbf{400ms}$, regardless of increasing dimensionality. The non-caching agents, however, ran into a mean runtime of $\mathbf{721ms}$ for 3 dimensions. Also, (2) caching did show no sign of compromising the SLO fulfillment; on the contrary, from Fig.~\ref{subfig:E3_1_SLO_F} we observe that it stabilized the SLO fulfillment towards more satisfying assignments: for 3 dimensions, the caching agent outperformed the non-caching agent by $\mathbf{32\%}$ SLO fulfillment. 

\vspace{5pt}
\subsubsection{Scalability of RASK -- Processing Services (\textbf{E6})}

The number of processing services naturally dictates the complexity of optimizing their parameter assignments. Suppose our approach works scale-free, it could ensure equal SLO fulfillment for a larger number of services under analog resource restrictions. \changed{We evaluate the RASK agent in 3 different environments: the default setup with 3 services and maximum cores $c_{\text{max}} =8$; then 6 services with $c_{\text{max}} =16$; and finally 9 services with $c_{\text{max}} =24$.} Hence, the maximum number of available \textit{cores} grows proportionally to the services.

To avoid implementing six additional service types, we replicate the existing three services (i.e., QR, CV, and PC) and spawn up to three containers for each image. However, from an agents perspective, different instances of the same service are treated completely independent at inference time. Hence, the agent uses the same regression function for up to three QR services, but infers independent parameter assignments for each. Thus, the solver has to optimize $|P| = \{7, 14, 21\}$ parameters for $|S| =\{3, 6, 9\}$ services respectively. Again, we evaluate the agent under the \textit{Diurnal} pattern and stabilize results by running 5 repetitions per instance.

\paragraph{Results}

\begin{figure}[t]
    \subfloat[Runtime Complexity]{%
    \includegraphics[width=0.499\columnwidth]{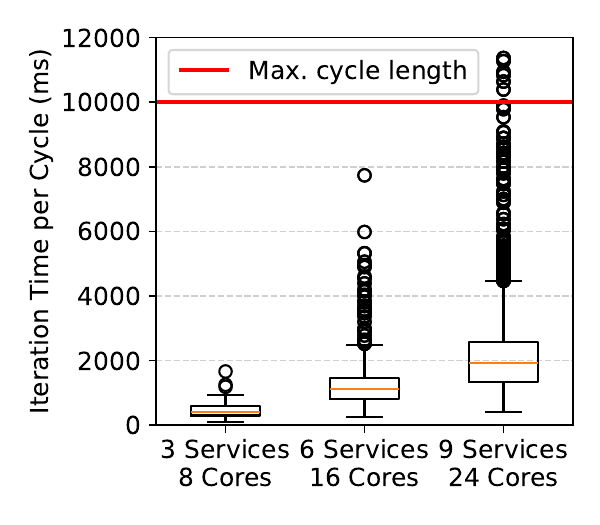}%
    \label{subfig:E3_2_complexity}%
    }%
    \subfloat[SLO Fulfillment]{%
        \includegraphics[width=0.499\columnwidth]{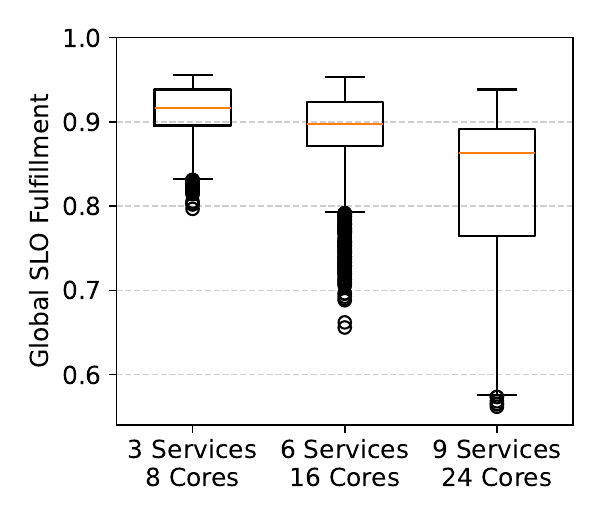}%
    \label{subfig:E3_2_SLO_F}%
    }
    \caption{Runtime complexity and SLO fulfillment of RASK for increasing numbers of processing services under equivalent resource constraints. Agent faces \textit{Diurnal} request pattern.}
    \label{fig:E3_2_Performance}%

\end{figure}

Fig.~\ref{fig:E3_2_Performance} shows again the distributions of the agent's runtime (ms) and the respective global SLO fulfillment---this time averaging up to 9 services. We notice that: (1) the runtime increases linearly with the numbers of services, leading to a median runtime of $\mathbf{2s}$ for 9 services; however, even when caching the last parameter assignment, RASK occasionally (i.e., in the outliers) showed spikes of runtimes larger than $\mathbf{10s}$---thus delaying the autoscaling interval. Also, (2) despite the larger resource constraints, the SLO fulfillment slightly declined with increasing numbers of services: 9 services produced a median fulfillment of $\mathbf{0.87}$ over all 5 runs. We will carefully interpret this behavior in Section~\ref{sec:discussion}, where we discuss the natural limitations of the numerical solver and potential strategies to solve this.

\section{Discussion}
\label{sec:discussion}

In the following, we extract a series of implications from the presented results and use them to answer the posed challenges.

\changed{
$\textit{C1:}$
Our results support a shift to multi-dimensional elasticity: \textbf{E4} showed how controlling more elasticity dimensions increases the SLO fulfillment of RASK almost linearly (cfr. Fig.~\ref{subfig:E3_1_SLO_F_Cache}). RASK achieved this by analyzing the impact of parameter interventions on the service performance (cfr. Fig.~\ref{fig:regression-functions}); also, the SLO weights (cfr. Table~\ref{tab:slo-thresholds}) allowed customizing how the agent builds tradeoffs between parameters.}
Meanwhile, RASK's operational overhead was low: $\mathbf{0.043}$ CPU cores for optimizing 9 services (cfr. Table~\ref{tab:rask-overhead}).
To decrease RASK's runtime latency, we observe that caching the last result accelerates the numerical solver significantly, while simultaneously improving SLO fulfillment (cfr. Fig.~\ref{fig:E3_1_Performance}).
Still, when analyzing the scalability in \textbf{E6}, we observe that higher numbers of services ($|S| \geq 9$) face increased runtime and SLO violations under proportional resource constraints (cfr. Fig.~\ref{fig:E3_2_Performance}). This shows the natural limitations of the numerical solver used, to which we see numerous solutions:
(1) detecting similarities (e.g., profiles) between services---in our case we assumed completely independent services; 
\changed{(2) offloading the autoscaling task to a remote node with more resources;
(3) parallelizing the autoscaling task between multiple agents to split overall complexity; 
(4) taking advantage of RASK's containerization and allocate more resources;
and (5) decreasing the model training frequency or run in a detached thread---Table~\ref{tab:latency-breakdown} showed that this dominated overall latency.}

\implication{Multi-dimensional autoscaling improves the SLO fulfillment on resource-constrained Edge devices with increasing dimensionality. This creates a complex optimization problem that requires a sophisticated solver.}

$\textit{C2:}$
Our findings underline that RASK can quickly build service-specific scaling policies: \textbf{E1} showed that 20 iterations in the environment---corresponding to $\mathbf{200s}$ runtime---were sufficient to develop a generalizable model for three heterogeneous processing services (cfr. Fig.~\ref{fig:E1_SLO_F}). During this time, the RASK agent explores each services' action space randomly; this might be accelerated by avoiding visited configurations or assigning a dedicated information gain~\cite{sedlak_equilibrium_2024} to certain areas in the configuration space. This also avoids rare cases (i.e., one run for \textbf{E1} $\{\xi =20, \eta=0\}$) where the random exploration did not find satisfying parameter assignments. After this short training period, the agent was fully trained for solving the other experiments---including request pattern that it had not experienced before.
%
%
To increase the accuracy of the regression models, \textbf{E2} also showed how a custom polynomial degree per service can decrease the MSE significantly: using the recommended polynomial degree $\delta = 4$ for the QR service reduces the MSE by a factor of \textbf{2.4} (cfr. Table~\ref{tab:degree-mse}). RASK required $\mathbf{395ms}$ for adjusting 3 services in up to 3 dimensions (cfr. Fig.~\ref{fig:E3_1_Performance_Cache}), 
which includes training the regression models and running the numerical solver. Under these circumstances, we see high benefit and little overhead from learning or adjusting the optimal polynomial degree during service runtime.

\implication{Using few samples, RASK develops an accurate model of the processing environment that supports service-specific scaling policies. Efficient exploration and refinement are needed to ensure model accuracy.}

$\textit{C3:}$
Our results highlight RASK's superiority against SOTA under resource scarcity: \textbf{E3} showed that RASK sustains dynamic request pattern with up to 28\% higher SLO fulfillment compared to a default k8s VPA and a RL-based autoscaler (cfr. Fig.~\ref{fig:E2_SLO_F}). The performance gap between RASK and the baselines was particularly large during peak load, when RASK was the only service to sustain the SLO fulfillment of the CV service by tuning its qualitative parameters. 
All the evaluated agents operated on our multi-dimensional autoscaling platform, thus benefiting from vertical scaling without any cold start. However, only the RASK agent operated in a continuous action space (cfr. Fig.~\ref{fig:regression-functions}), which highlights its strengths for fine-grained scaling of CPU quota or data quality. While a potential caveat of our method is the increased development effort for exposing service-specific parameters, we believe that the increased SLO fulfillment pays this off.


\implication{Compared to SOTA, multi-dimensional autoscaling achieves higher SLO fulfillment during periods of high request load. Continuous scaling actions support fine-grained adjustments of services and their resources.}

\section{Conclusion}
\label{sec:conclusion}
In this work, we introduced MUDAP, a Multi-dimensional Autoscaling Platform tailored for resource-constrained Edge environments, along with RASK, a regression-based scaling agent that enables fine-grained vertical scaling of both resource-level and service-level parameters. Our approach addresses key challenges in modern Edge computing—namely, ensuring SLO fulfillment under strict resource constraints, tailoring elasticity strategies to heterogeneous services, and adapting to dynamic request patterns.
We evaluate our approach using three stream processing services: a Yolov10 object detector, a QR code reader, and a Lidar renderer.
Our results show that RASK can learn service-specific autoscaling models within less than 20 autoscaling cycles (i.e., $200s$ of processing).
Compared to Kubernetes VPA and DQN-based baselines, our method achieves up to 28\% fewer SLO violations, particularly during periods of high load, while introducing minimal resource overhead. Most notably, we demonstrated how an increasing number of elasticity dimensions leads to higher SLO fulfillment.
%
These contributions pave the way for more responsive and efficient autoscaling mechanisms that go beyond traditional resource-focused strategies.





\section*{Acknowledgment}

This work was partially supported by the European Union's Horizon Europe research and innovation program under grant agreements No. 101135576 (INTEND) and No. 101070186 (TEADAL). Also, this work was partially supported by CNS2023-144359 and the European Union NextGenerationEU/PRTR under MICIU/AEI/10.13039/501100011033.

\bibliographystyle{IEEEtran}
\bibliography{Boris,references}

\begin{IEEEbiography}[{\includegraphics[width=1in,height=1.25in,clip,keepaspectratio]{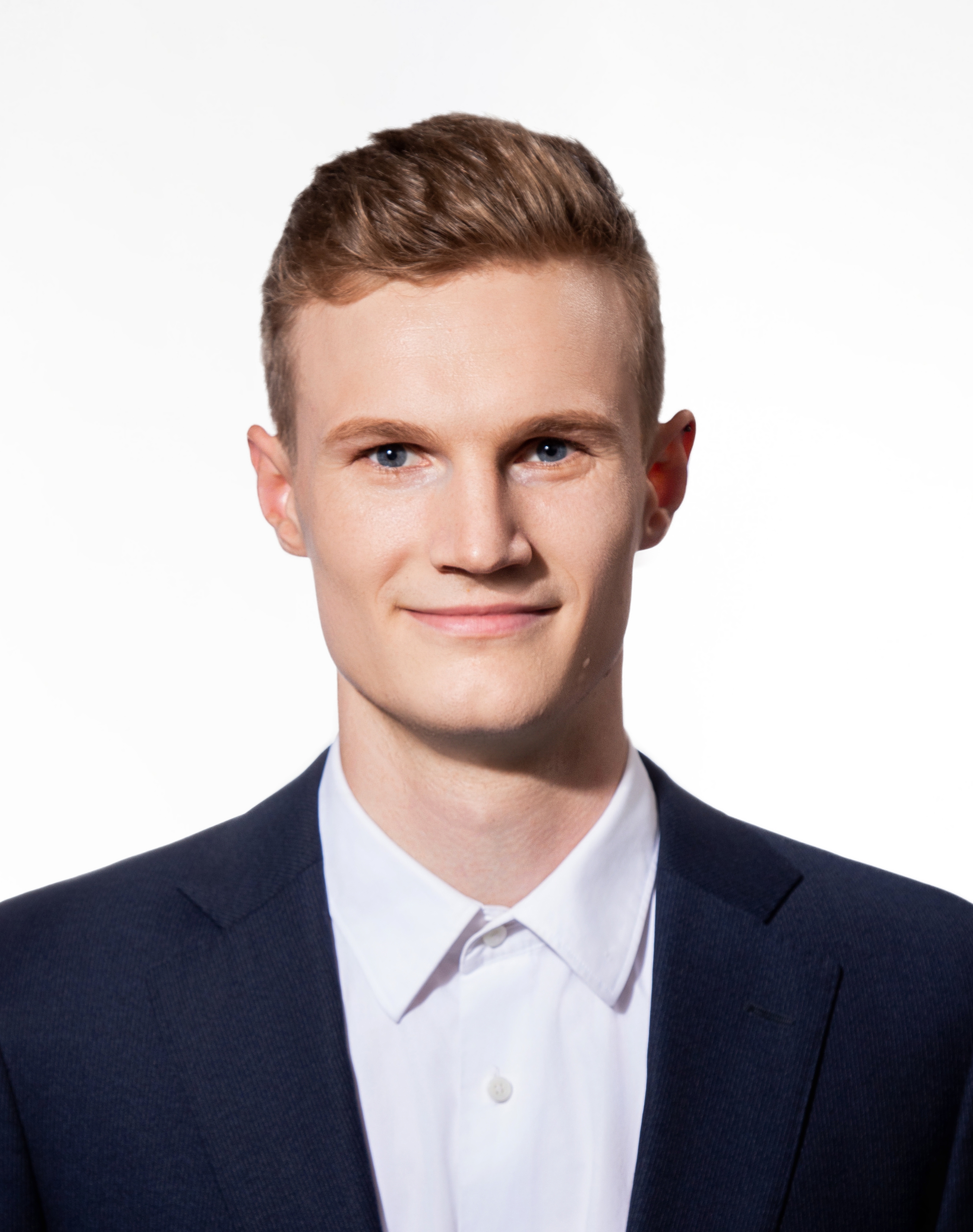}}]{Boris Sedlak} is a postdoctoral researcher at the Distributed Systems Group at TU Wien, Austria, where he received his PhD in 2025. Prior to that, he received his B.Sc. in Media Informatics at the University of Applied Sciences in St. P\"olten, and his M.Sc. in Software Engineering \& Internet Computing at the TU Wien. His research interests include edge intelligence, causal methods for the computing continuum, and service-oriented computing.
\end{IEEEbiography}

\begin{IEEEbiography}[{\includegraphics[width=1in,height=1.25in,clip,keepaspectratio]{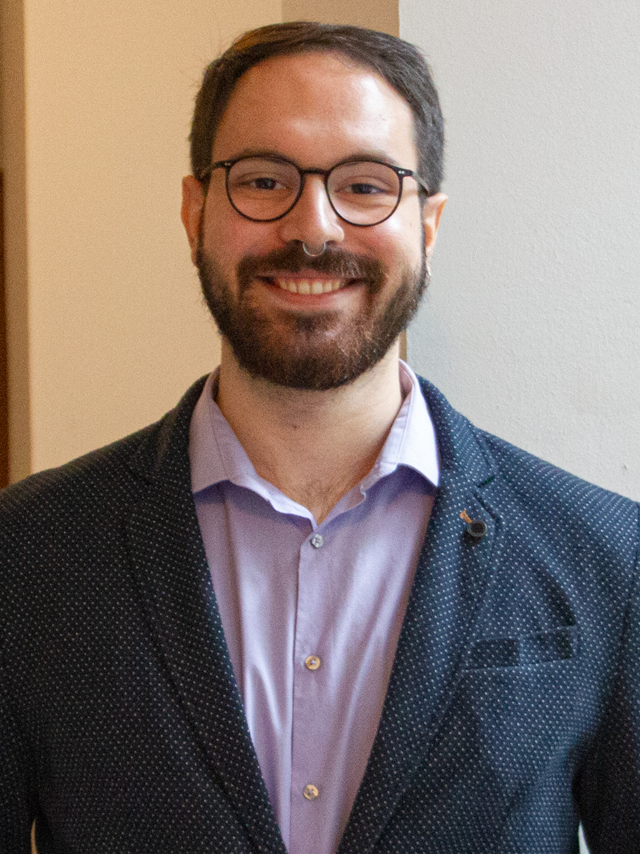}}]{Philipp Raith} received a MSc from the Technical University of Vienna, Austria in 2021 with distinction in the field of Computer Science. He is now a PhD candidate at the Distributed Systems Group in the field of Edge Computing. His research interests include Serverless Edge Computing, Edge Intelligence and Operations for AI.
\end{IEEEbiography}

\begin{IEEEbiography}[{\includegraphics[width=1in,height=1.25in,clip,keepaspectratio]{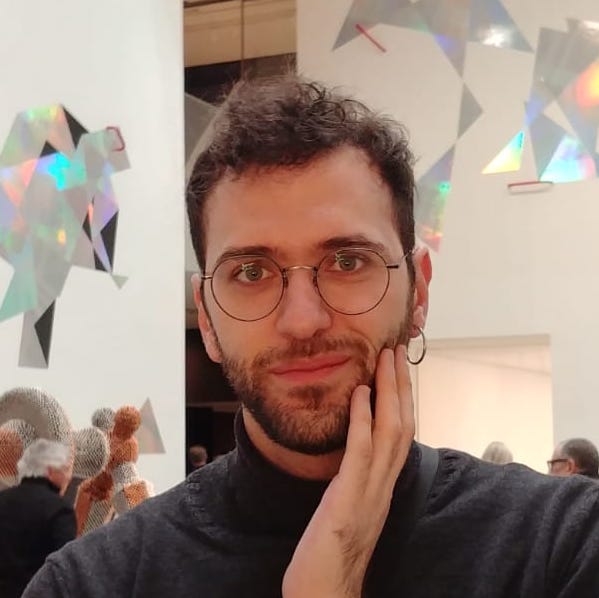}}]{Andrea Morichetta} is a postdoctoral researcher at the Distributed Systems Group of TU Wien, specializing in machine learning in and for distributed systems, and edge-to-cloud computing. He holds a Ph.D. in Electrical, Electronics, and Communication Engineering from Politecnico di Torino. He has collaborated with leading institutions and industry partners such as Cisco (San Jose, CA, US), and Tsinghua University (Beijing, CN).\end{IEEEbiography}

\begin{IEEEbiography}[{\includegraphics[width=1in,height=1.25in,clip,keepaspectratio]{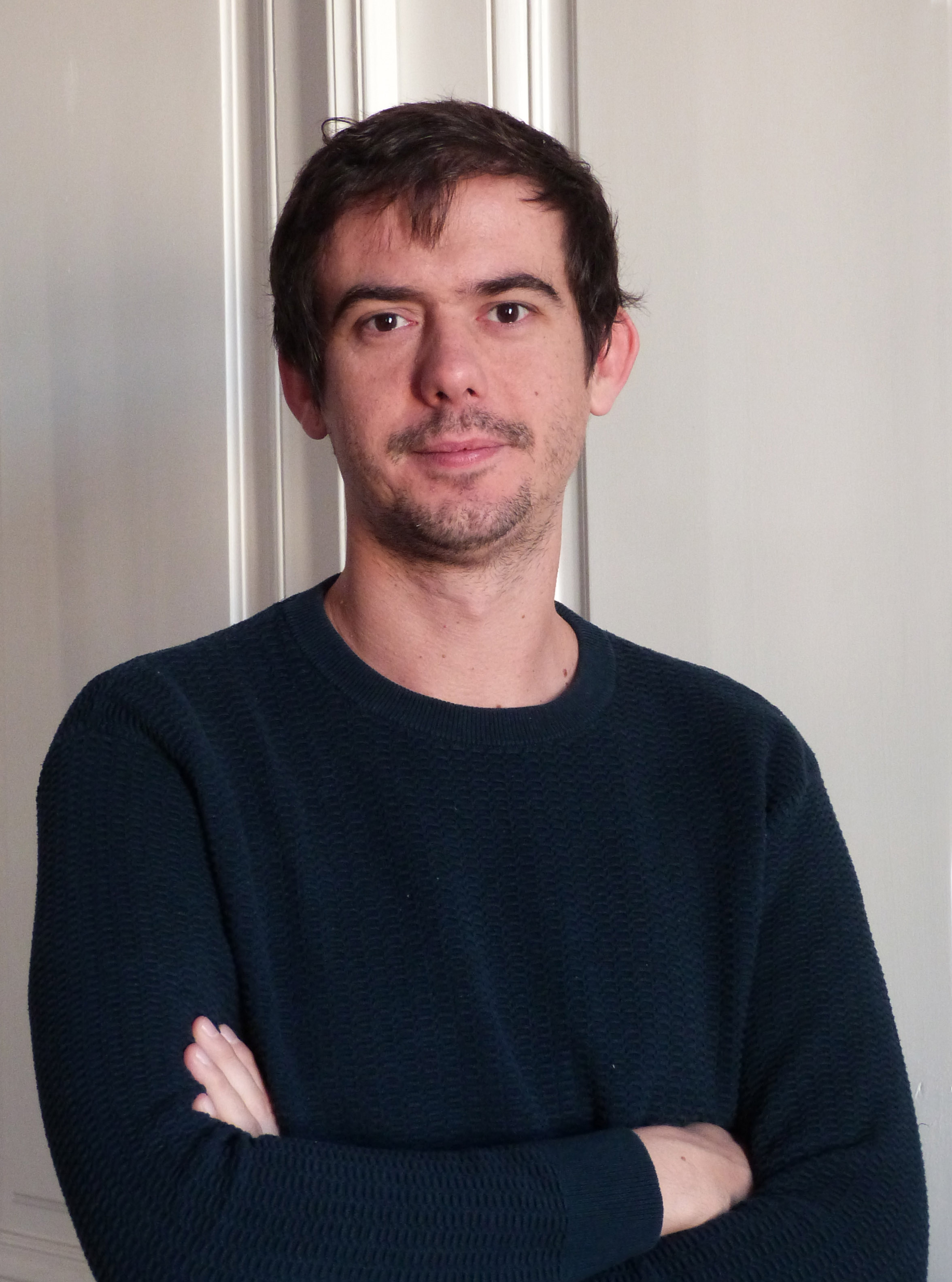}}]{Víctor Casamayor Pujol}
is a tenure-track assistant professor at Universitat Pompeu Fabra (UPF), Barcelona, Spain. Prior to that, he worked as a postdoctoral researcher in the Distributed Systems Group at TU Vienna after having earned his PhD at UPF. His research interests revolve around self-adaptive methodologies for computing continuum systems, including SLO-based definitions, causal and machine learning inference, and robotics. 
\end{IEEEbiography}

\begin{IEEEbiography}[{\includegraphics[width=1in,height=1.25in,clip,keepaspectratio]{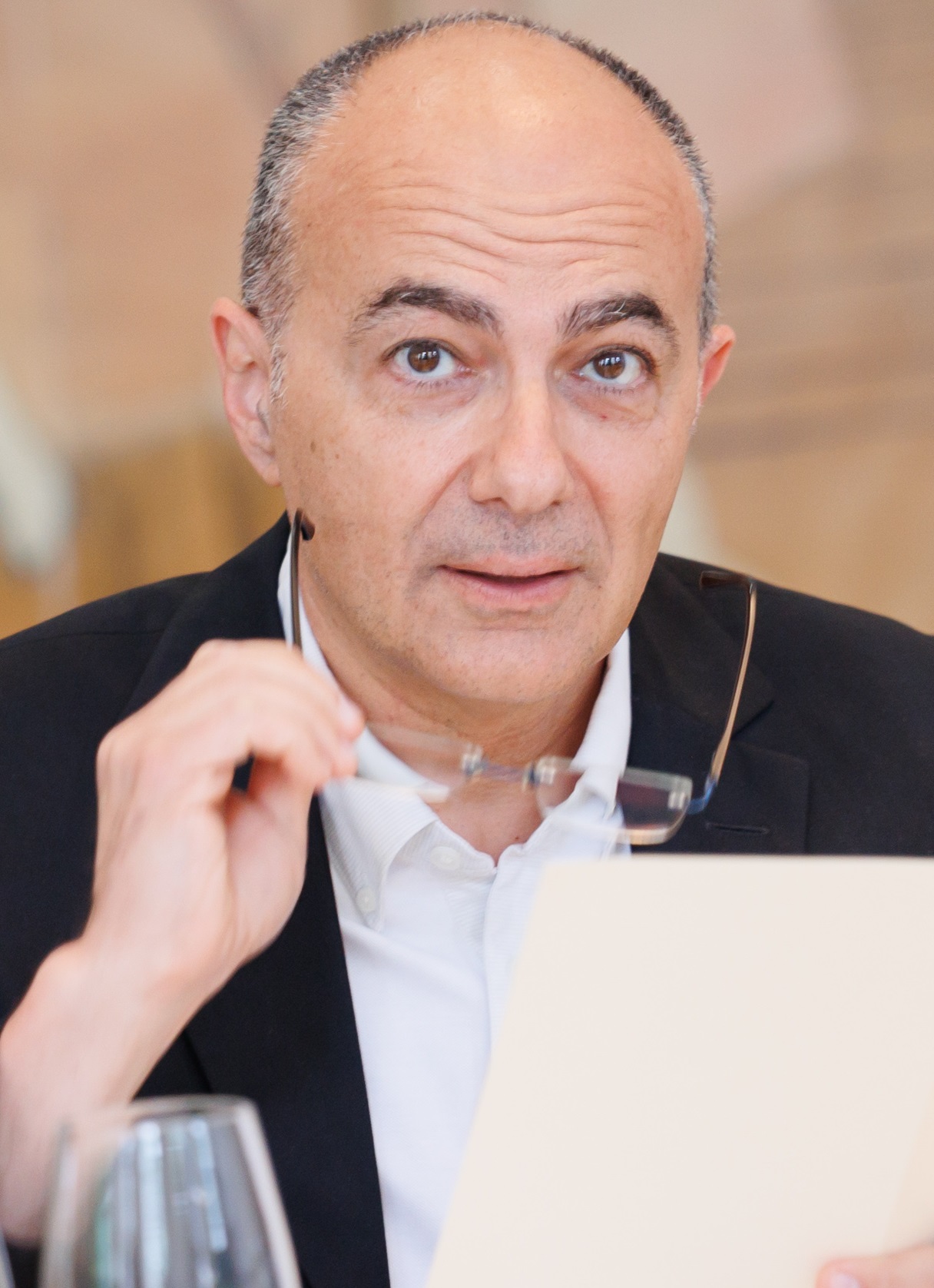}}]{Schahram Dustdar}
received his Ph.D. in business informatics from the University of Linz, Austria, in 1992. He is a full professor of computer science, focusing on internet technologies, and heads the Distributed Systems Group at TU Wien, Vienna, Austria. He is chairman of the Informatics Section of the Academia Europaea. He has been an associate editor for IEEE Transactions on Services Computing, ACM Transactions on the Web, and ACM Transactions on Internet Technology. He has received several accolades, including the ACM Distinguished Scientist and IBM Faculty awards.
\end{IEEEbiography}

\end{document}